\documentclass[12pt]{article}
\usepackage{amsmath}
\usepackage{a4wide}
\usepackage{amssymb}
\usepackage{graphicx}
\begin{document}
{\renewcommand{\thefootnote}{\fnsymbol{footnote}}
\hfill  IGPG--07/6--2\\
\medskip
\begin{center}
{\LARGE  Effective equations for isotropic quantum cosmology including
matter}\\
\vspace{1.5em}
Martin Bojowald\footnote{e-mail address: {\tt bojowald@gravity.psu.edu}}
\\
\vspace{0.5em}
Institute for Gravitational Physics and Geometry,
The Pennsylvania State
University,\\
104 Davey Lab, University Park, PA 16802, USA\\
\vspace{0.5em}
Hector Hern\'andez\footnote{e-mail address: {\tt hhernandez@uach.mx}}
\\
\vspace{0.5em}
Universidad Aut\'onoma de Chihuahua,
Facultad de Ingenier\'ia,\\
Nuevo Campus Universitario, Chihuahua 31125, M\'exico\\
\vspace{0.5em}
Aureliano Skirzewski\footnote{e-mail address: {\tt askirz@gmail.com}}
\\
\vspace{0.5em}
Centro de F\'isica Fundamental,
Universidad de los Andes, M\'erida 5101, Venezuela
\vspace{1.5em}
\end{center}
}

\setcounter{footnote}{0}

\newcommand{\lP}{\ell_{\mathrm P}}

\newcommand{\md}{{\mathrm{d}}}
\newcommand{\tr}{\mathop{\mathrm{tr}}}
\newcommand{\sgn}{\mathop{\mathrm{sgn}}}

\newcommand*{\R}{{\mathbb R}}
\newcommand*{\N}{{\mathbb N}}
\newcommand*{\Z}{{\mathbb Z}}
\newcommand*{\Q}{{\mathbb Q}}
\newcommand*{\C}{{\mathbb C}}

\newcommand{\be}{\begin{equation}}
\newcommand{\ee}{\end{equation}}
\newcommand{\bq}{\begin{eqnarray}}
\newcommand{\eq}{\end{eqnarray}}

\begin{abstract}
  Effective equations often provide powerful tools to develop a
  systematic understanding of detailed properties of a quantum system.
  This is especially helpful in quantum cosmology where several
  conceptual and technical difficulties associated with the full
  quantum equations can be avoided in this way. Here, effective
  equations for Wheeler--DeWitt and loop quantizations of spatially
  flat, isotropic cosmological models sourced by a massive or
  interacting scalar are derived and studied. The resulting systems are
  remarkably different from that given for a free, massless
  scalar. This has implications for the coherence of evolving states
  and the realization of a bounce in loop quantum cosmology.
\end{abstract}

\section{Introduction}

While initial stages of the expanding branch of our universe are
becoming better and better understood, its potential history before
the big bang remains open to many debates. By now, several scenarios
have been suggested which, building on quantum gravity effects rather
than purely classical gravity, could give rise to a non-singular
evolution at all times
\cite{CyclicEkpy,Emergent,Oscill,Cyclic,EmergentLoop,svv}. The era
preceding the classical big bang singularity, as well as the quantum
transition through it, can then be relevant to explain observed
features of our universe and must therefore be thoroughly
understood. Most of these scenarios realize the simplest possible
non-singular behavior where an isotropic collapsing universe reaches a
smallest, non-zero volume before it bounces back into an expanding
branch. In addition to isotropy, one often assumes a special form of
matter, such as a free massless scalar making quantum cosmological
models solvable. Whether or not this results in a robust, reliable
scenario for a complex universe can only be seen by embedding such
simple models in a more general class, at least perturbatively,
including matter interactions or inhomogeneities. From a conceptual
viewpoint, this provides a stability analysis of the original model;
from a phenomenological viewpoint it will allow one to construct
realistic scenarios with all ingredients relevant for observations.

Loop quantum cosmology gives rise to a potentially general mechanism
which removes space-time singularities
\cite{Sing,SphSymmSing,BSCG}. In this general form, not much is said
about the precise regime around the big bang and the state before, and
especially at, the classical singularity might have been highly
quantum. This, in fact, agrees with general expectations.  It thus
came as quite a surprise when detailed properties derived for an
isotropic model in loop quantum cosmology sourced by a free, massless
scalar indicated that the state remained semiclassical to a high
degree throughout its evolution, before and after the big bang
\cite{QuantumBigBang}. This model was analyzed because the absence of
a potential allows a direct calculation of observables and an
implementation of the physical inner product. This, in turn, made it
possible to evaluate numerical solutions to the underlying evolution
equations for physical information \cite{APS,APSII}. When such special
mathematical simplifications, beyond those already implied by assuming
space-time symmetries, are used it is never clear whether detailed
properties are general or specific to the chosen model. This
especially applies to the semiclassical aspects which were unexpected
from the point of view of general quantum systems whose states usually
spread. An analytical analysis of a related and exactly solvable
model\footnote{Solvability requires one to drop some of the
ingredients of a loop quantized Hamiltonian which, however, are not
relevant for states of a universe with large
matter content. The solvable model is thus not exactly the same as a
loop quantization or that used in the numerical studies of \cite{APS}
but still allows one to analyze the bounce. Note that also in
\cite{APS} the dynamical equation was adapted for the numerical
purposes and differs from what one would obtain in a loop quantization
\cite{LivRev}. None of these changes matter for properties of the
bounce of a large universe.}  \cite{BouncePert,BounceCohStates}
clarified this issue and showed that the semiclassicality properties
were indeed very special: the bounce is described by a solvable model
in which expectation values and fluctuations (or higher moments of a
wave function) do not couple to each other. Thus, while the state
evolves, its expectation values are not affected by its spreading, and
the spreading is completely independent of other parameters of the
state.  In other words, the system is analogous to a harmonic
oscillator in quantum mechanics whose states can remain coherent
forever while following the classical trajectory. As quantum systems
go, this high degree of solvability is a rare property well-known from
the harmonic oscillator or free quantum field theories but not
corresponding to a general feature.

In this model, expectation values, spread parameters and higher
moments of a state evolve independently of each other and can be
determined exactly. Thus, while the wave function evolves and possibly
spreads and deforms, this does not influence the trajectory followed
by its expectation values.  But when details of the model are changed,
solvability can no longer be maintained and complicated coupling terms
between expectation values, spreads and higher moments arise. This
leads to quantum corrections to the classical equations which can
often be captured in effective equations
\cite{EffAc,EffectiveEOM,Karpacz}. In this paper, we set out to study
such effective equations for an isotropic model in loop quantum
gravity with a {\em massive and self-interacting} scalar in order to
test the genericness of properties of the exactly solvable model.
While the solvability of the exact bounce model certainly makes its
properties very special, it also provides the starting point for a
systematic perturbation analysis. This is analogous to free field
theories used as the zeroth order in perturbative quantum field
theory. Adding a potential as done here is the simplest generalization
in that it does not introduce additional degrees of freedom. (Initial
steps to include inhomogeneities have been done in
\cite{QuantCorrPert}.)  However, it also leads to difficulties since
the scalar, instead of being a monotonic function of coordinate time
as in the free case, has itself non-trivial dynamics. As discussed
in more detail below, a monotonic scalar is used crucially in the
solvable model since evolution of the universe volume is described
relationally with respect to the scalar.  If the scalar is no longer
monotonic, it cannot be used as a global time coordinate, meaning that
the model with a potential can only be used for a finite range of
time. Nevertheless, we will see that this allows one to test whether
quantum back-reaction effects remain negligible or whether they can
have a strong effect on the behavior of a bounce.

It is often assumed that a potential term of a scalar field cannot
have a sizable effect near an isotropic classical singularity because
its classical form $a^3V(\phi)$ tends to zero for $a\to 0$ while the
kinetic term $\frac{1}{2}a^{-3}p_{\phi}^2$ in canonical variables has
a diverging pre-factor. The general behavior certainly depends also on
how $p_{\phi}$ and $\phi$ behave when the singularity is approached,
but if the potential is small $p_{\phi}$ does not change much, being a
constant of motion in the free case. Thus, the kinetic term is
expected to dominate and a non-zero potential should not change much
of the bounce observed in free models. However, this reasoning
overlooks the quantum nature of the problem. It is based on a
classical Hamiltonian and the evolution equations it implies. If
quantum gravity is used to argue for the free bounce, quantum aspects
must also be taken into account for interactions implied by the
potential. Here, details of the free bounce related to solvability
become important. The solvable bounce model demonstrates that a state
which starts out semiclassically at large volume and evolves to
smaller volume stays semiclassical and then, {\em in this
semiclassical form,} bounces when quantum geometry effects set in. The
model also shows that there do exist states, though not semiclassical
ones, which do not bounce \cite{BounceCohStates}. This does not alter the
conclusions about singularity removal in the free model since one
always has to use the boundary condition that a state is semiclassical
at large volume (at least at one side of the bounce). Under this
condition, every solution of the free model bounces. But this is no
longer guaranteed with interacting matter. Any realistic statement
about singularity removal has to involve evolution over long stretches
of time: a universe which is semiclassical at large volume, like our
own, must be shown to remain non-singular after a long time of
backward (or forward) evolution.  This does not play a large role for
a solvable model since there are no quantum back-reaction effects of
an evolving state on its expectation values. But a non-zero potential
introduces such back-reaction effects which must be considered even if
they are small for a perturbative potential. A reliable statement
about non-singular behavior must involve a proof that a state starting
out semiclassically at large volume does not spread and deform too
much so as to avoid a bounce, which may well be possible since the
free model allows non-semiclassical states which do not bounce. In
general, the bounce regime, or whatever regime is valid around the
classical big bang singularity, is expected to be of a highly quantum
nature; arguments based solely on states which are semiclassical in
this regime, by choice or by the lack of quantum back-reaction in a
specific model, cannot be complete.

In this paper, we provide effective equations taking into account
quantum back-reaction in the presence of a perturbative potential and
provide an initial investigation.  This is a first step of a stability
analysis of bounce models in loop quantum cosmology. We will find
several new properties and arrive at a set of effective equations of a
new type requiring the presence of independent quantum degrees of
freedom. Even though by construction we need to require the potential
to be small and a perturbation to the kinetic term, several new
qualitative features can be seen. The analytical, though approximate,
analysis allows more general conclusions than numerical investigations,
which are often prone to hidden assumptions, would show.

\section{Description of the models}

We start with the Friedmann equation in an isotropic and homogeneous
background with a scalar field $\phi$. It can be written as a
constraint $C=0$ with
\begin{equation} \label{Friedmann}
C=-\frac{3}{8\pi G}c^2 \sqrt{|p|}+ 
\frac{1}{2}|p|^{-3/2} p_{\phi}^2 + |p|^{3/2} V(\phi)\,,
\end{equation}
in terms of canonical variables $|p|=a^2$, $c=\md a/\md\tau$ (using
proper time $\tau$) for the metric and $p_{\phi}=a^3 \md\phi/ \md\tau$
as the canonical momentum conjugate to $\phi$. We will be ignoring
factors of $8\pi G/3$ in this article to keep subsequent formulas
simple. A convenient understanding of dimensions then is to use
dimensionless $c$ and $\phi$ while $p$ and $1/V$ have the dimension of
length squared. Also the Planck constant $\hbar$ has the dimension of
length squared if $G$ is ignored in $\hbar G=\ell_{\rm P}^2$ with the
Planck length $\ell_{\rm P}$. The variable $p$ is derived from an
isotropic densitized triad \cite{IsoCosmo} and can thus in general
assume both signs to take into account the triad orientation. In this
paper, however, it will be sufficient to use only positive values of
$p$ since our aim is to analyze bounces; a change of sign in $p$ would
imply that no bounce occurs. (Even if quantum effects imply regular
evolution through the classical singularity at $p=0$, this would not
be considered a bounce which conventionally requires a non-zero lower
bound to volume. ``Bounce'' is usually also reserved for situations in
which a semiclassical description remains valid at all times and the
universe does not enter a deep quantum regime. This is realized in all
examples for bounces described so far.)

The Friedmann equation describes the dynamics of an isotropic
space-time
\[
 \md s^2=-\md\tau^2+a(\tau)^2 (\md x^2+\md y^2+\md z^2)
\]
with the scale factor $a$, sourced by a homogeneous scalar matter
field with potential $V(\phi)$. Since (\ref{Friedmann}) is a
constraint, it generates equations of motion by $\md f/\md t=\{f,NC\}$
for any phase space function $f$, referring to a derivative by a time
coordinate $t$ which corresponds to the chosen lapse function $N$. For
proper time $\tau$, as in the above line element, we simply have
$N=1$. Then,
\begin{eqnarray}
 \md p/\md \tau &=& 2\sqrt{|p|}c\\
 \md c/\md \tau &=& -\frac{c^2}{2\sqrt{|p|}}-
\frac{3}{2|p|}\left(\frac{p_{\phi}^2}{2|p|^{3/2}}-
|p|^3V(\phi)\right)\\
 \md\phi/\md \tau &=& |p|^{-3/2}p_{\phi}\\
 \md p_{\phi}/\md \tau &=& -|p|^{3/2}V'(\phi)\,.
\end{eqnarray}
The first and third equation determine the relation between
momenta and time derivatives of coordinates, while the other two then
result in second order evolution equations for $p$ and $\phi$ in
proper time.

\subsection{Internal time}

Instead of using a coordinate time $\tau$, we can solve
(\ref{Friedmann}) for $p_{\phi}$,
\begin{equation}
 \frac{1}{\sqrt{2}}|p_{\phi}|= |p| \sqrt{ c^2-|p| V(\phi)}=:H
\end{equation}
and view the right hand side as a Hamiltonian generating relational
evolution in $\phi$. Rather than solving for $p(\tau)$, $c(\tau)$ and
$\phi(\tau)$ where, due to general covariance the choice of time
coordinate $\tau$ does not matter, we thus eliminate coordinate time
altogether and look for solutions $p(\phi)$ and $c(\phi)$.  We will
deal here mainly with a perturbative scheme considering the potential
to be small in comparison with the gravitational contribution. Thus,
we will often refer to the expanded Hamiltonian
\begin{equation}
H= |p| c-\frac{p^2}{2c} V(\phi) +O(V^2)
\end{equation}
up to second order in $V$. 

Relational solutions in terms of $\phi$ can easily be found explicitly
in the potential free case since $p_{\phi}$ is a constant and thus
$\phi$ is a monotonic function of $\tau$. Any point in $\tau$ is
mapped one-to-one to a value of $\phi$. However, if there is a
non-trivial potential such as a mass term $\frac{1}{2}m^2\phi^2$, in
general $\phi$ will no longer be monotonic but shows oscillations
around minima of the potential. Only stretches between two turning
points of $\phi$ can then be described in an internal time form, but
not all of the evolution that would be accessible in $\tau$. Moreover,
the system is more complicated due to an explicit ``time'' dependence
from $V(\phi)$ appearing in $H$. As we will see, upon quantization
this still allows us to investigate the effects of spreading and
deforming wave packets on the expectation values of a corresponding
quantum system. In particular, interesting phases in early universe
cosmology are inflationary which, if sourced by an inflaton, typically
provides a slow-roll phase where the scalar rolls down a flat
potential. In fact, we will see shortly that the potential we deal
with perturbatively must be flat so as to provide this setting
automatically. This implies a long monotonic phase for $\phi$ during
which back-reaction effects could build up and become important. In
backward evolution, this allows us to see if quantum effects could
possibly prevent or dramatically change a bounce.

\subsection{Wheeler--DeWitt quantization}

A further difficulty related to the potential arises at the quantum
level. To see this we first recall from \cite{BouncePert} how the
potential-free case is treated quantum theoretically. Instead of a
time dependent Hamiltonian, this case would give us simply $
\frac{1}{\sqrt{2}} |p_{\phi}| = |pc|$ which, when quantized, results
in a Schr\"odinger equation
\begin{equation} \label{first}
 i\hbar \frac{1}{\sqrt{2}} \frac{\partial}{\partial\phi}\psi=
\widehat{|pc|}\psi\,.
\end{equation}
Solutions to this equation automatically solve the equation
\begin{equation} \label{second}
 -\frac{\hbar^2}{2}\frac{\partial^2}{\partial\phi^2} \psi=
\widehat{p^2c^2}\psi
\end{equation}
which (in a certain factor ordering) quantizes (\ref{Friedmann})
before taking the square root. Since the Friedmann equation, from the
canonical point of view, is the Hamiltonian constraint, it is indeed
quantized first without solving for $p_{\phi}$ in canonical quantum
gravity. Although (\ref{second}) is a second order differential
equation and thus has more solutions than the first order equation
(\ref{first}), one usually allows superpositions only of solutions
with a fixed sign of energy $p_{\phi}$ which is chosen positive in
taking the root above. Thus, the solution space to (\ref{second})
relevant for the quantum theory should be considered as consisting of
two sectors, one of which is given by solutions to (\ref{first}) and
the other by solutions with the opposite sign of the square root.

A solvable model is then obtained by using the Hamiltonian
$\hat{H}=\frac{1}{2}(\hat{c}\hat{p}+\hat{p}\hat{c})$ for
$\phi$-evolution which is quadratic in canonical coordinates, just as
the harmonic oscillator Hamiltonian. This is responsible for the
solvability, a fact which, more surprisingly, even extends to the loop
quantization \cite{BouncePert} as recalled in the next
subsection. Solvability implies that the equations of motion (written
in internal time $\phi$ whose derivative is denoted by a dot)
\begin{equation} \label{ExpEOM}
 \langle\dot{\hat{p}}\rangle =
\frac{\langle[\hat{p},\hat{H}]\rangle}{i\hbar} =
-\langle\hat{c}\rangle \quad,\quad \langle\dot{\hat{c}}\rangle =
\frac{\langle[\hat{c},\hat{H}]\rangle}{i\hbar} =
\langle\hat{p}\rangle 
\end{equation}
form a closed system and can thus be solved easily without knowing the
complicated ways a state may spread and deform. In general quantum
systems, by contrast, such equations would form a system of infinitely
many coupled equations also involving the quantum variables
\begin{equation}
 G^{a,n}= \left\langle ((\hat{c}-\langle\hat{c}\rangle)^{n-a}
(\hat{p}-\langle\hat{p}\rangle)^a)_{\rm Weyl}\right\rangle
\end{equation}
for $2\leq n\in{\mathbb N}$ and $0\leq a\leq n$
\cite{EffAc,EffectiveEOM}, the subscript ``Weyl'' denoting fully
symmetric ordering. Strictly speaking, this is even the case in the
system considered here since we dropped the absolute value of the
Hamiltonian to make it quadratic. As discussed in detail in
\cite{BounceCohStates}, the quadratic constraint still allows one to
describe states correctly which are sharply peaked at large values of
energy $H$ or $p_{\phi}$, a case of primary interest for isotropic
cosmological models. For the quadratic Hamiltonian, explicit solutions
to (\ref{ExpEOM}) for expectation values $c=\langle\hat{c}\rangle$ and
$p=\langle\hat{p}\rangle$ as well as fluctuations are given by 
\begin{eqnarray}
 c_{V=0}(\phi) &=& C_1e^{\phi} \label{FreeWDWI}\\
 p_{V=0}(\phi) &=& C_2e^{-\phi} \label{FreeWDWII}\\
 G_{V=0}^{cc}(\phi) &=& C_3e^{2\phi} \label{FreeWDWIII}\\
 G_{V=0}^{cp}(\phi) &=& C_4 \label{FreeWDWIV}\\
 G_{V=0}^{pp}(\phi) &=& C_5e^{-2\phi} \label{FreeWDWV}
\end{eqnarray}
which we distinguish from the perturbative solutions with
non-vanishing potential determined later by the subscript.  Since to
the orders analyzed here we are mainly interested in fluctuations, we
will use only quantum variables at $n=2$, and call them
$G^{cc}:=G^{0,2}$, $G^{cp}:=G^{1,2}$ and $G^{pp}:= G^{2,2}$ for better
clarity.

In the presence of a potential we would, following the same procedure,
first quantize to obtain a second order equation
\begin{equation} \label{secondpot}
  -\frac{\hbar^2}{2}\frac{\partial^2}{\partial\phi^2} \psi=
\left(\widehat{p^2c^2}-|\hat{p}|^3V(\phi)\right)\psi\,.
\end{equation}
Now, however, $[\hat{p}_{\phi},\hat{H}]\not=0$ for a quantization of
\begin{equation}
 H= |p|\sqrt{c^2-|p|V(\phi)}\,.
\end{equation}
Thus, solutions to
\begin{equation} \label{Schroedinger}
 -\frac{1}{\sqrt{2}}\hat{p}_{\phi}\psi=\hat{H}\psi
\end{equation}
do not solve the second order equation (\ref{secondpot}) but rather
\begin{equation} \label{pphisquared}
 \frac{1}{2}\hat{p}_{\phi}^2\psi=\hat{p}_{\phi}\hat{H}\psi=
\hat{H}\hat{p}_{\phi}\psi+ [\hat{p}_{\phi},\hat{H}]\psi= \hat{H}^2\psi
+ [\hat{p}_{\phi},\hat{H}]\psi\,.
\end{equation}
There is thus no strict analog of the first order equation,
complicating any analysis.

However, for the solutions we are interested in here, which includes
perturbations around solutions of the exact free scalar model
obtained for large $H$, we can ignore the commutator term. In fact,
due to the square root in $\hat{H}$ we have, up to factor ordering in
a precise quantization,
\[
 [\hat{p}_{\phi},\hat{H}] \sim i\hbar\frac{|\hat{p}^3|
V'(\phi)}{2\hat{H}}
\]
whose expectation value is small compared to that of $\hat{H}^2$, also
appearing in (\ref{pphisquared}) for states with large $H$ and a not
too steep potential. This is exactly the regime we intend to probe by
our perturbation theory. Such a regime easily arises without
fine-tuning and can thus provide general insights: The two conditions
$|p|^3V(\phi)\ll p^2c^2 \approx H^2$ for the potential term to be
perturbative and $\hbar p^3V'(\phi)/H\ll H^2$ for the commutator in
(\ref{pphisquared}) to be negligible require $V\ll H^2/p^3$ and $\hbar
V'\ll H^3/p^3$. In the loop quantized model to be discussed next, a
semiclassical bounce in the free model occurs for $H\gg\hbar$ for the
universe not to enter the deep quantum regime, and we have $p\geq
H$. Thus, around such a bounce the conditions on the potential are
$\hbar V\ll 1$ and $\hbar V'\ll 1$, such that both the potential and
its first derivative must be small compared to the Planck mass
squared. For a sufficiently small and flat potential, long evolution
times can be considered. For the purpose of describing those regimes,
the first order equation (\ref{Schroedinger}) is a good approximation,
and due to large $p_{\phi}$ we can again drop absolute values on $H$.

\subsection{Loop quantization}

It is more interesting to study perturbations around the solvable
model determined by a loop Hamiltonian following again
\cite{BouncePert} for the zeroth order solutions. In a loop
quantization there is no operator for $c$ \cite{Bohr}, and one can
thus not use a Hamiltonian of the same form as before. Instead, all
exponentials $e^{i\alpha c}$ with $\alpha\in{\mathbb R}$ are
represented by well-defined operators and appear in the Hamiltonian
such that the classical expression $cp$ is reproduced in the
classical, small-curvature limit. This is accomplished by the free
Hamiltonian $p
\sin{c}$ as it follows from the loop quantization (up to quantization
choices which we can ignore here). This Hamiltonian is no longer
quadratic in the canonical variables and one could thus expect strong
quantum back-reaction even in the free model. However, this turns out
not to be the case since one can choose non-canonical variables $p$
together with $J=pe^{ic}$ in which the system becomes linear: We have
classical Poisson relations
\begin{equation}
 \{p,J\}_{\rm class} = -iJ\quad,\quad \{p,\bar{J}\}_{\rm class} =
 i\bar{J} \quad,\quad
\{J,\bar{J}\}_{\rm class} = 2ip\,.
\end{equation}
Moreover, the Hamiltonian $H=\frac{1}{2i}(J-\bar{J})$ is a linear
combination of the basic variables, making the dynamical system
$(p,J,H)$ a linear one. For such systems, if they remain linear after
quantization, no quantum back-reaction occurs. (A quadratic
Hamiltonian in canonical variables is a special case of a linear
system.) That the system does remain linear after quantization has
been demonstrated in \cite{BouncePert}.

In the presence of a potential the Hamiltonian becomes\footnote{As
written, this refers to exponentials depending only on connection
components while recent investigations of loop quantum cosmology have
suggested the use of triad dependent holonomies, i.e.\ $\alpha(p)$ in
$e^{i\alpha c}$, in Hamiltonian constraint operators
\cite{APSII,InhomLattice,SchwarzN}. While basic holonomies in the
holonomy-flux algebra do not depend on triad components, the
appearance of triad dependent holonomies can be motivated by lattice
refinements of an inhomogeneous state occurring in a physical
state. Our following discussion of the qualitative behavior does not
depend much on which form of holonomies is used in the constraint
since alternative cases can be mapped into each other by canonical
transformations of $(p,J)$ before quantization.}
\begin{equation} \label{HLoop}
H= \sqrt{\left( \frac{J- \bar{J}}{2i}\right)^2- p^3 V(\phi)},
\end{equation}
which we again treat perturbatively.  The expansion in $V$ reads
\begin{equation}
H=  \frac{J-\bar{J}}{2i}-i \frac{p^3}{J-\bar{J}}V(\phi)+ O(V^2(\phi)).
\end{equation}

After quantization, we again obtain an equation of the form
(\ref{Schroedinger}). In the variables used here, $(p,J)$, we can use
loop techniques and quantize $J$ using holonomies. If we choose the
ordering $\hat{J}=\hat{p}\widehat{e^{ic}}$, basic commutators are
\begin{equation}
 [\hat{p},\hat{J}]=\hbar\hat{J} \quad,\quad
[\hat{p},\hat{J}^{\dagger}]=-\hbar\hat{J}^{\dagger} \quad,\quad
[\hat{J},\hat{J}^{\dagger}]= -2\hbar(p+\hbar/2)\,.
\end{equation}
Quantum Poisson brackets obtained by taking expectation values of
these equations are thus corrected compared to the classical ones:
\begin{equation}
 \{p,J\} = -iJ\quad,\quad \{p,\bar{J}\} =
 i\bar{J} \quad,\quad
\{J,\bar{J}\} = 2i(p+\hbar/2)\,.
\end{equation}
Moreover, in these variables there are more than three fluctuation
parameters, namely $G^{pp}$, $G^{J\bar{J}}$, $G^{pJ}$, $G^{p\bar{J}}$,
$G^{JJ}$ and $G^{\bar{J}\bar{J}}$, since we are using one complex
variable $J$. These quantum variables are defined analogously as with
canonical variables, e.g.\
\[
 G^{J\bar{J}} = \frac{1}{2}\langle\hat{J}\hat{J}^{\dagger}
+\hat{J}^{\dagger}\hat{J}\rangle -J\bar{J}\,.
\]
They are restricted to the correct number by reality
conditions $G^{p\bar{J}}= \overline{G^{pJ}}$,
$G^{\bar{J}\bar{J}}=\overline{G^{JJ}}$ as well as
\begin{equation} \label{reality}
 |J|^2-(p+{\textstyle\frac{1}{2}}\hbar)^2=G^{pp}-G^{J\bar{J}}+
 \frac{1}{4}\hbar^2=  \frac{1}{4}\hbar^2-c_1\,.
\end{equation}
as a consequence of
$\hat{J}\hat{J}^{\dagger}=\hat{p}^2$. ($G^{J\bar{J}}-G^{pp}=:c_1$
turns out to be a constant of motion for the free model. Due to its
appearance in the reality condition it plays an important role which
will also be seen in the perturbed models discussed in what follows.)

Dynamically, the difference to the Wheeler--DeWitt case is
that $\hat{H}$ in a metric or triad representation is a difference
rather than differential operator \cite{cosmoIV,IsoCosmo}. Instead of
a partial differential equation in the presence of a scalar we then
have a partial difference-differential equation (see also
\cite{Scalar,ScalarLorentz}). This is recognizable in our treatment
only indirectly since we only implicitly deal with constraint
equations for states. We rather express the information contained in
the equation for a state through equations implied for expectation
values, fluctuations and, if needed, higher moments of the state. The
main effect then is the presence of $J$ or $e^{ic}$, not $c$ itself,
in equations of motion. Also effective equations will then initially
depend on $e^{ic}$ instead of $c$, which suggests that they are simply
obtained by a replacement of $c$ by $\sin c$ as the real quantity
having $c$ as the low curvature limit for $c\ll 1$. However, this
replacement on its own overlooks coupling terms of fluctuations which
will be analyzed in this paper to provide a complete set of effective
equations.

The replacement of $c$ by $\sin c$ is complete for the free model as
shown in \cite{BouncePert} and indicated by the numerical studies of
\cite{QuantumBigBang,APS}. In this case, the quantum Hamiltonian lacks
coupling terms from fluctuations and free solutions 
\begin{eqnarray}
 p_{V=0}(\phi) &=& \frac{1}{2}(Ae^{-\phi}+Be^{\phi})-\frac{1}{2}\hbar
 \label{FreeLoopI} \\
 J_{V=0}(\phi) &=&
\frac{1}{2}(Ae^{-\phi}-Be^{\phi})+iH_0\label{FreeLoopII}
\end{eqnarray}
for expectation values and 
\begin{eqnarray}
 G_{V=0}^{pp}(\phi) &=& \frac{1}{2}(c_3e^{-2\phi}+c_4e^{2\phi})- 
\frac{1}{4}(c_1+c_3)\label{FreeLoopIII} \\
 G_{V=0}^{JJ}(\phi) &=& \frac{1}{2}(c_3e^{-2\phi}+c_4e^{2\phi})+ 
\frac{1}{2}(3c_2-c_1)-
2i(c_5e^{\phi}-c_6e^{-\phi})=
\overline{G_{V=0}^{\bar{J}\bar{J}}(\phi)}\label{FreeLoopIV} \\
 G_{V=0}^{pJ}(\phi) &=& \frac{1}{2}(c_3e^{-2\phi}-c_4e^{2\phi})+ 
i(c_5e^{\phi}+c_6e^{-\phi}) =
\overline{G_{V=0}^{p\bar{J}}(\phi)}\label{FreeLoopV} \\
G_{V=0}^{J\bar{J}}(\phi) &=& \frac{1}{2}(c_3e^{-2\phi}+c_4e^{2\phi})+
\frac{1}{4}(3c_1-c_2)\,.\label{FreeLoopVI} 
\end{eqnarray}
for all fluctuations can be determined explicitly.

\section{Effective equations}

Hamiltonian operators determine the evolution equation
(\ref{Schroedinger}) for a wave function $\psi$. Equivalently, a state
can be described in terms of its moments, or its expectation values of
basic operators together with quantum variables. As already used in
(\ref{ExpEOM}) and seen in detail later, a Hamiltonian operator can
then also be used to derive equations of motion for these infinitely
many variables directly, without taking the detour of a wave
function. These infinitely many equations are in general coupled and
difficult to analyze. Effective equations provide a well-defined
scheme to control these equations by formulating them as equations of
motion for the classical variables $c$ and $p$ identified with
expectation values, or for a larger but finite set including also some
of the quantum variables. Such equations can be obtained by systematic
approximations (mainly the semiclassical one) to neglect most of the
moments, or by solving for all but finitely many of the quantum
variables {\em in terms of those variables to be retained} and
inserting those solutions into the equations of motion for the
variables to be retained. In the simplest case, one thus obtains
effective equations for $c$ and $p$ amended by terms such as an
effective potential depending on $c$ and $p$. This is different from
other methods such as perturbation schemes where solutions are already
obtained as functions of time; one would solve all the equations up to
a certain order but {\em including those for $c$ and $p$} to find
solutions. Solving for or eliminating only quantum variables but
keeping a finite set of variables free requires different techniques.
The result of such a procedure, a set of effective equations, will be
much more powerful than perturbative solutions since it will
illustrate quantum effects more directly and more generally by
equations of motion (e.g.\ containing an effective potential) rather
than only through specific solutions.

In general, the derivation of effective equations is more complicated
than perturbative solution procedures, and it is not even guaranteed
that effective equations of a given type exist to describe a certain
regime of interest well. The key step in their derivation is a
decoupling of the infinitely many equations for moments of a quantum
system such that almost all of them can be solved at least
approximately (or shown to be negligible) in terms of a finite set of
remaining variables. The main tool to achieve this is often an
adiabatic expansion for quantum variables which, for instance,
provides the low energy effective action for anharmonic oscillators in
quantum mechanics. But quantum variables such as fluctuations need not
behave adiabatically in a given regime, or only certain combinations
may behave so. Thus, in general the procedure of effective equations
is not guaranteed to be of success. As we will see now, quantum
cosmology provides examples for regimes where effective equations of
adiabatic type do not exist. Nevertheless, as we will also show,
bounces in loop quantum cosmology can, fortunately, be described by
effective equations even though they turn out to be rather involved.

\subsection{Effective equations in quantum cosmology}

As shown in \cite{BouncePert,BounceCohStates} and recalled in the
preceding section, for a specific ordering of Hamiltonian constraint
operators the equations of motion (\ref{ExpEOM}) for expectation
values of the triad $p$ and connection $c$ describing an isotropic
universe sourced by a free, massless scalar do not depend on any
quantum variables. They can not only be taken immediately as effective
equations without genuine quantum corrections but even describe the
system exactly. This is analogous to the equations governing
expectation values of a harmonic oscillator or free field theory
state. It is true not only in the Wheeler--DeWitt quantization but
even in a loop quantization of the same model in appropriate
variables. The replacement of connection components $c$ by
``holonomies'' $\sin c$ in the constraint and the Friedmann equation,
as often done as a shortcut to loop effective equations, is thus
strictly justified for these models. In the presence of a potential,
however, the situation is more complicated and additional quantum
corrections arise from the back-reaction of fluctuations on the
behavior of expectation values. These corrections can be crucial and
must be determined for a robust analysis of quantum systems. This is
analogous to adding an anharmonic contribution to the harmonic
oscillator, or interaction terms to a free quantum field theory.

In the latter, well-known cases an adiabatic approximation can be
employed usefully to describe the quantum behavior around the ground
or vacuum state of the system. Some of these corrections take into
account properties of the {\em interacting vacuum state} as opposed to
the free vacuum, or of {\em dynamical coherent states} as opposed to
just kinematical ones. Quantum corrections can be computed order by
order in an expansion in $\hbar$ combined with the adiabatic
one. While the $\hbar$-expansion is always of interest to capture
quantum effects in semiclassical regimes, an adiabatic expansion has
to be justified independently. For perturbations around ground states
of the harmonic oscillator or free field theories this is clearly
reasonable: The exact states (simple Gaussians) of the unperturbed
theories have constant fluctuations, and thus perturbations should
result only in a weak time dependence. This expectation is borne out
self-consistently by the fact that the adiabatic approximation even
with interaction terms provides well-defined solutions.

The situation is different, however, for the quantum cosmological
models studied here. We do have a solvable system which can provide
the zeroth order for such an expansion, but it lacks a ground state or
an otherwise distinguished state in which fluctuations and other
quantum variables would be constant. Instead, exact solutions
(\ref{FreeWDWI})--(\ref{FreeWDWV}) for the Wheeler--DeWitt model and
(\ref{FreeLoopI})--(\ref{FreeLoopVI}) for the loop quantization in the
absence of interactions show that most quantum variables are
exponential functions of time. It is thus not clear a priori which
combination of quantum variables, if any, can be treated
adiabatically. Nevertheless, any candidate can be analyzed
self-consistently by trying to set up an adiabatic approximation. As
we will see in the descriptions that follow, several subtleties arise
in this process. It clearly demonstrates that for general systems in
loop quantum cosmology there is much more to an effective analysis
than a simple replacement of connection components $c$ by $\sin c$. We
will discuss this issue further in our Conclusions section, after
providing a detailed analysis of effective descriptions.

\subsection{The Wheeler--DeWitt model}

We start with an attempt to develop effective equations based on an
adiabaticity assumption for quantum variables. While this will
ultimately be unsuccessful, it illustrates the main procedure and
several difficulties which can arise. After that we describe possible
choices of other combinations of the variables which may be assumed to
be adiabatic, although none of them will result in well-defined
effective equations for the Wheeler--DeWitt model. In view of a
comparison with the loop model, and to confirm that the analyzed cases
do not lead to consistent effective equations, it is nevertheless
instructive to go through some of the details. In the next section, we
will study a different, higher-dimensional type of effective equations
which does exist.

To derive effective equations from a Hamiltonian operator $\hat{H}$ one
first writes the expectation value of $\hat{H}$ in a general state as
a functional of expectation values, spreads and higher moments,
captured in quantum variables
\begin{equation}
 G^{a,n}= \left\langle ((\hat{c}-\langle\hat{c}\rangle)^{n-a}
(\hat{p}-\langle\hat{p}\rangle)^a)_{\rm Weyl}\right\rangle
\end{equation}
for $2\leq n\in{\mathbb Z}$ and $0\leq a\leq n$. This is possible
because, with certain restrictions such as uncertainty relations
discussed below, the quantum variables together with expectation
values allow one to reconstruct the state $\psi(c,p,G^{a,n})$ which
has been used in computing the expectation value of $\hat{H}$. Thus,
the quantum Hamiltonian $H_Q(c,p,G^{a,n})=
\langle\psi(c,p,G^{a,n})|\hat{H}|\psi(c,p,G^{a,n})\rangle$ is a
function of expectation values $c$ and $p$ and quantum variables
$G^{a,n}$. Due to the square root present in $H$ it is difficult to
write an explicit Hamiltonian operator in our case. Fortunately, this
can be circumvented in our framework where we do not work with states
and operators directly but with the coupled dynamics of expectation
values and quantum variables. We thus implicitly define a Hamiltonian
operator, order by order in a perturbation expansion as follows.

The quantum Hamiltonian can be obtained by formally expanding
\begin{eqnarray}
 H_Q= \langle H(\hat{c},\hat{p})\rangle &=& \langle
H(c+(\hat{c}-c),p+(\hat{p}-p))\rangle\nonumber\\
 &=& \sum_{n=0}^{\infty} \sum_{a=0}^n
\frac{1}{n!}\binom{n}{a}\frac{\partial^n
H(c,p)}{\partial p^a\partial c^{n-a}}G^{a,n} \label{HQGen}
\end{eqnarray}
in $\hat{c}-c$ and $\hat{p}-p$ (with $G^{a,1}$ understood as taking
the value one, i.e.\ we use normalization of states; for $n=1$ we have
$G^{a,1}=0$ by definition). Higher order terms in these expansion
parameters then combine to correction terms to the classical
Hamiltonian containing quantum variables:
\begin{equation}\label{HQWdW}
 H_Q = cp -\frac{1}{2}\frac{p^2}{c} V(\phi)- \frac{p^2}{2c^3}
 V(\phi)G^{cc}+ \left( 1+\frac{p}{c^2}V(\phi) \right)G^{cp} 
-\frac{1}{2c}V(\phi) G^{pp} + \cdots
\end{equation}
where we restrict ourselves to positive $p$ from now on. Since we
treat $\phi$ as an internal time variable, we do not include quantum
variables of $\phi$ and $p_{\phi}$ or any correlations between matter
and metric degrees of freedom. We do, however, include fluctuations
and correlations of the gravitational variables relevant to the given
orders, which is second in moments and first in the potential. Note
that this suggests correction terms in an effective potential of the
form
\begin{equation} \label{Veff}
 V_{\rm eff}(\phi) = (1+c^{-2}G^{cc}(\phi)- 
2c^{-1}p^{-1}G^{cp}(\phi)+ p^{-2}G^{pp}(\phi))V(\phi)
\sim (1+C_3-2C_4 +C_5)V(\phi)
\end{equation}
(if we use the free solutions in the second step) and a ``zero point
energy'' $G^{cp}\sim C_4$, which is independent of the expectation
values, to the Hamiltonian. To use this for an effective Hamiltonian
in terms of $c$ and $p$ only, we would have to solve consistently for
the quantum variables by appropriate means. A precise realization, as
we will see, turns out to be more complicated.

Equations of motion for expectation values and quantum variables are
derived from quantum commutators determining Poisson brackets
between expectation values and all the quantum variables. For
instance, we have $\{c,p\}=\langle[\hat{c},\hat{p}]\rangle/i\hbar=1$
and, using the Leibniz rule,
\begin{eqnarray}
 \{G^{pp},G^{cp}\} &=& \{\langle\hat{p}^2\rangle-
\langle\hat{p}\rangle^2, {\textstyle\frac{1}{2}}\langle\hat{c}\hat{p}+
\hat{p}\hat{c}\rangle- \langle\hat{c}\rangle\langle\hat{p}\rangle\}=
-2G^{pp}\label{FluctPoissonI}\\
 \{G^{cc},G^{cp}\} &=& 2G^{cc} \label{FluctPoissonII}\\
 \{G^{cc},G^{pp}\} &=& 4G^{cp} \label{FluctPoissonIII}\,.
\end{eqnarray}
General formulas are provided in \cite{EffAc}. Equations of motion on
this infinite dimensional phase space derived from the quantum
Hamiltonian $H_Q(c,p,G^{a,n})=\langle\hat{H}\rangle$ agree with the
quantum evolution equations determined from commutators with the
Hamiltonian operator.  From the quantum Hamiltonian (\ref{HQWdW}) we
then obtain equations of motion
\begin{eqnarray}
 \dot{c} &=& \{c,H_Q\} = p-\frac{p}{c} \left(1+c^{-2}G^{cc}
-(cp)^{-1}G^{cp}\right)V(\phi)\,,\nonumber\\
 \dot{p} &=& \{p,H_Q\} = -c-\frac{1}{2} \frac{p^2}{c^2}
\left(1+\frac{3}{2} c^{-2}G^{cc}-
4(cp)^{-1}G^{cp}+\frac{1}{2}p^{-2}G^{pp}\right)V(\phi)\,,\\
 \dot{G}^{a,2} &=& \{G^{a,2},H_Q\}= a\frac{p^2}{c^3}V(\phi) G^{a-1,2}+ 2(1-a)
\left(1+\frac{p}{c^2}V(\phi)\right) G^{a,2}- (2-a)\frac{1}{c}V(\phi)
G^{a+1,2}\,. \nonumber
\end{eqnarray}

\subsubsection{Non-adiabaticity of quantum variables}

To analyze possible adiabaticity of fluctuations, we use their
equations of motion
\begin{eqnarray}
 \dot{G}^{cc} &=& 2\left(1+\frac{p}{c^2}V(\phi)\right) G_0^{cc}-
\frac{2}{c}V(\phi) G_0^{cp}\,, \\
 \dot{G}^{cp} &=& \frac{p^2}{c^3}V(\phi) G_0^{cc}- \frac{1}{c}V(\phi)
G_0^{pp}\,,\\
 \dot{G}^{pp} &=& 2\frac{p^2}{c^3}V(\phi) G_0^{cp}- 2
\left(1+\frac{p}{c^2}V(\phi)\right) G_0^{pp}\,.
\end{eqnarray}
These will be the only equations needed for the
order we are interested in. An adiabatic approximation proceeds by
expanding the fluctuations as $G^{a,2}=
\sum_{e=0}^{\infty}\lambda^eG_e^{a,2}$, replacing $\md/\md \phi$ by
$\lambda\md/\md \phi$ in the equations of motion and expanding the
resulting equations in $\lambda$. One can then solve order by order
for $G_e^{a,2}$ and sum the contributions. The adiabatic approximation
to a given order $N$ is the sum $G_N^{a,2}=
(\sum_{e=0}^N\lambda^eG_e^{a,2})|_{\lambda=1}$. As the example of the
low energy effective action for anharmonic oscillators shows, the
zeroth order is already sufficient to derive an effective potential,
and quantum corrections to the mass will be obtained at second order
\cite{EffAc,EffectiveEOM}.

Proceeding in this way, we obtain equations
\begin{eqnarray}
 0 &=& 2\left(1+\frac{p}{c^2}V(\phi)\right) G_0^{cc}- \frac{2}{c}V(\phi)
 G_0^{cp}\,, \label{ZerothGWDWI}\\
 0 &=& \frac{p^2}{c^3}V(\phi) G_0^{cc}- \frac{1}{c}V(\phi)
G_0^{pp}\,,\label{ZerothGWDWII}\\
 0 &=& 2\frac{p^2}{c^3}V(\phi) G_0^{cp}- 2
\left(1+\frac{p}{c^2}V(\phi)\right) G_0^{pp}\label{ZerothGWDWIII}
\end{eqnarray}
solved by
\begin{eqnarray}
 G_0^{cp} &=& \frac{c}{V(\phi)}
 \left(1+\frac{p}{c^2}V(\phi)\right) G_0^{cc}\,, \label{Gcpzero}\\
 G_0^{pp} &=& \frac{p^2}{c^2} G_0^{cc}
\end{eqnarray}
in terms of $G_0^{cc}$. Only these two relations follow since
(\ref{ZerothGWDWI})--(\ref{ZerothGWDWIII}) presents a degenerate
system of linear equations. This is a general property in these
variables since the equations are derived with the degenerate Poisson
algebra (\ref{FluctPoissonI})--(\ref{FluctPoissonIII}) of three
fluctuation variables.  The free variable $G_0^{cc}$, which in general
can be a function on phase space and of $\phi$, has to be determined
at least up to a finite number of constants which specify the state
used to expand around. Such a restriction arises by considering the
next order: At first adiabatic order we have equations
\begin{eqnarray}
 \dot{G}_0^{cc} &=& 2\left(1+\frac{p}{c^2}V(\phi)\right) G_1^{cc}-
 \frac{2}{c}V(\phi) G_1^{cp}\,,\\ 
\dot{G}_0^{cp} &=&
 \frac{p^2}{c^3}V(\phi) G_1^{cc}- \frac{1}{c}V(\phi) G_1^{pp}\,,\\
 \dot{G}_0^{pp} &=& 2\frac{p^2}{c^3}V(\phi) G_1^{cp}- 2
\left(1+\frac{p}{c^2}V(\phi)\right) G_1^{pp}
\end{eqnarray}
which is again a degenerate linear system, but now an inhomogeneous
one for $G_1^{a,2}$. This can only be solved for the $G_1^{a,2}$
provided that the relation
\begin{eqnarray}
 0&=& \frac{p^2}{c^2}\dot{G}_0^{cc}+
\dot{G}_0^{pp}-2\frac{c}{V(\phi)}\left(1+\frac{p}{c^2}V(\phi)\right)
\dot{G}_0^{cp}\nonumber\\
&=&\frac{p^2}{c^2} \dot{G}_0^{cc}+ \frac{\md}{\md\phi}
\left(\frac{p^2}{c^2}G_0^{cc}\right)-
2\left(\frac{p}{c}+\frac{c}{V(\phi)}\right)
\frac{\md}{\md\phi}\left( \left(\frac{p}{c}+\frac{c}{V(\phi)}\right)
G_0^{cc}\right)
\end{eqnarray}
is satisfied, a differential equation which in fact
determines $G_0^{cc}$ up to a constant. One can solve this equation
for arbitrary potential $V(\phi)$ by
\begin{equation}
 G_0^{cc}= \frac{xV(\phi)}{\sqrt{c^2+2pV(\phi)}}
\end{equation}
with the constant of integration $x$, and thus
\begin{eqnarray}
 G_0^{cp} &=&
\frac{xc(1+pV(\phi)/c^2)}{\sqrt{c^2+2pV(\phi)}}\,,\\
 G_0^{pp} &=& \frac{xp^2V(\phi)/c^2}{\sqrt{c^2+2pV(\phi)}}\,.
\end{eqnarray}
Note that the initially worrisome inverse $V(\phi)$ in (\ref{Gcpzero})
drops out of the final expressions which are perturbative in
$V(\phi)$. Nevertheless, having to divide by the potential in an
intermediate step will turn out to be problematic. This feature will
be cured in the loop quantized model.

These solutions can now be inserted into the equations of motion for
$c$ and $p$ to obtain candidate effective equations to zeroth
adiabatic order
\begin{eqnarray}
 \dot{c}&=&
c-\frac{p}{c}\left(1-\frac{x}{p\sqrt{c^2+2pV(\phi)}}\right) V(\phi)\nonumber\\
&\sim& c-\frac{p}{c}\left(1-\frac{x}{cp}+ \frac{x}{c^3}V(\phi)\right)
V(\phi) + O((pV/c^2)^3)\label{ceff}\\
\dot{p}&=& -p+-\frac{p^2}{2c^2}
\left(1-\frac{4x}{p\sqrt{c^2+2pV(\phi)}}\right) V(\phi)\nonumber\\
& \sim& -p-
\frac{p^2}{2c^2} \left(1-\frac{4x}{cp}+
\frac{4x}{c^3}V(\phi)\right)V(\phi) +O((pV/c^2)^3)\label{peff}
\end{eqnarray}
There is no $\hbar$ in these effective equations, but the free
variable $x$ parameterizes fluctuations of a state. It should thus be
expected to be of the order $\hbar$ for effective equations
corresponding to states saturating the uncertainty relation
\begin{equation}
 G^{cc}G^{pp}-(G^{cp})^2\geq \frac{\hbar^2}{4}\,.
\end{equation}
However, when trying to implement these relations, we recognize that
the effective equations are inconsistent because there is in fact no
state corresponding to this evolution: uncertainty relations are not
satisfied.  Inserting our solutions to zeroth order in the
adiabatic expansion, we obtain
\[
 G_0^{cc}G_0^{pp}-(G_0^{cp})^2= -\frac{c^2}{V(\phi)^2}
\left(1+2\frac{pV(\phi)}{c^2}\right)
(G_0^{cc})^2<0
\]
for $pV\ll c^2$. Thus, there is no choice for the only free parameter
$x$ in $G_0^{cc}$ which would make the adiabatic solution consistent
with the uncertainty relation. This illustrates how one can
self-consistently test whether an assumption of adiabaticity for
certain variables results in reliable effective equations. Although
the adiabaticity assumption for quantum variables has not resulted in
consistent effective equations, Eqs.~(\ref{ceff}) and (\ref{peff})
suggest that true effective equations are more involved than just
implying an effective potential (\ref{Veff}): There are new
expectation value dependent coefficients, and higher powers of the
classical potential. In what follows we will look at other
adiabaticity assumptions, before turning to higher dimensional
effective equations in the next section.

\subsubsection{Non-perturbative potential}

Before changing our adiabaticity assumption, we probe whether the
perturbative treatment of the potential may be responsible for the
non-existence of effective equations. There could, after all, be
regimes in which the potential is important in making quantum
variables adiabatic. We thus start from a non-expanded classical
Hamiltonian $H=p\sqrt{c^2-pV(\phi)}$ and obtain the quantum
Hamiltonian
\begin{eqnarray}
 H_Q&=& p\sqrt{c^2-pV(\phi)} -
\frac{p^2V(\phi)}{2(c^2-pV(\phi))^{3/2}}G^{cc}+
\frac{c^3-cpV(\phi)/2}{(c^2-pV(\phi))^{3/2}} G^{cp}\\
&&-\frac{c^2-3pV(\phi)/4}{2(c^2-pV(\phi))^{3/2}} V(\phi) G^{pp}+
\cdots \nonumber
\end{eqnarray}
up to second order in quantum variables. Now, the zeroth order
adiabatic solutions take the form
\begin{eqnarray}
 G_0^{cp} &=& \frac{c}{V(\phi)}
 \frac{c^2-pV(\phi)/2}{c^2-3pV(\phi)/4} G_0^{cc}\\
 G_0^{pp} &=& \frac{p^2}{c^2-3pV(\phi)/4}G_0^{cc}
\end{eqnarray}
and the uncertainty product
\[
 G_0^{cc}G_0^{pp}-(G_0^{cp})^2= -\frac{c^2}{V(\phi)^2}
\frac{1-\frac{p}{c^2}V(\phi)-
\frac{3}{4}\left(\frac{p}{c^2}V(\phi)\right)^2+ \frac{3}{4}
\left(\frac{p}{c^2}V(\phi)\right)^3}{\left(1 -
\frac{3}{4}\frac{p}{c^2}V(\phi)\right)} (G_0^{cc})^2
\]
is still negative independently of the potential in the allowed range
$|pV(\phi)|/c^2<1$. (Allowing larger potentials would give only a
narrow range $1<pV(\phi)/c^2< 2/\sqrt{3}\approx 1.155$ where the
previous expression would be positive. But then no real $p_{\phi}$ is
possible. Thus, dropping the assumption of a perturbative potential
does not, by itself, give rise to consistent effective equations.

\subsubsection{Non-adiabaticity relative to expectation values}

While none of the quantum variables in
(\ref{FreeWDWI})--(\ref{FreeWDWV}) is constant for a given state of
the free Wheeler--DeWitt model which one may choose to perturb around,
relative fluctuations $G^{cc}/c^2$ and $G^{pp}/p^2$ are constant, and
similarly for other quantum variables. One may thus try an adiabatic
approximation for those combinations rather than one directly in terms
of quantum variables which fails, as we saw. Let us now assume that
variables of the form
\begin{equation}
 g^{a,n}:= c^{-n+a+l}p^{-a+l}G^{a,n}
\end{equation}
behave adiabatically where $l$ is some fixed integer to be chosen. For
the free solutions, the $g^{a,n}$ will be constant for any $l$ since
$cp$ is constant, and thus adiabaticity seems more likely. (The
constants $C_I$ in the free solutions are not necessary to keep in the
definition of $g^{a,n}$ since they can be absorbed without changing
adiabaticity properties.)

Equations of motion for the new variables follow from the product
rule:
\begin{eqnarray*}
 \dot{g}^{cc} &=& \frac{pV(\phi)}{c (c^2-pV(\phi))^{3/2}}
\left(\frac{3}{2}\left((l+8/3)c^2-(l+2)pV(\phi)\right) g^{cc}-
2(c^2-3pV(\phi)/4)g^{cp}\right)\,,\\
 \dot{g}^{cp} &=& \frac{pV(\phi)}{c (c^2-pV(\phi))^{3/2}}
\left(c^2g^{cc}+ \frac{3}{2}(l+1)(c^2-pV(\phi))g^{cp}-
(c^2-3pV(\phi)/4)g^{pp}\right)\,,\\
 \dot{g}^{pp} &=& \frac{pV(\phi)}{c (c^2-pV(\phi))^{3/2}}
\left(2c^2g^{cp}+
\frac{3}{2}\left((l-3/2)c^2-lpV(\phi)\right)g^{pp}\right)
\end{eqnarray*}
where we ignore quadratic terms in $g^{a,n}$.  The zeroth order
adiabatic solutions are
\begin{eqnarray}
 g_0^{cp} &=& \frac{3}{4}
\frac{(l+8/3)c^2-(l+2)pV(\phi)}{c^2-3pV(\phi)/4} g_0^{cc}\,,\\
 g_0^{pp} &=& -\frac{c^2((l+8/3)c^2-(l+2)pV(\phi))}{(c^2-3pV(\phi)/4)
((l-2/3)c^2-lpV(\phi))} g_0^{cc}
\end{eqnarray}
which solves the remaining zeroth order equation only
for $l=0$, $l=1$ or $l=2$.

The case $l=0$, i.e.\ an assumption of adiabaticity for
$g^{cc}=G^{cc}/c^2$, $g^{cp}=G^{cp}/cp$ and $g^{pp}=G^{pp}/p^2$, gives
solutions $g_0^{cp}=2g_0^{cc}$ and $g_0^{pp}=4g_0^{cc}$ which in this
case are actually exact to any adiabatic order. However, the
uncertainty product $c^2p^2g^{cc}g^{pp}-(cpg^{cp})^2$ now vanishes for
any choice of free parameters and uncertainty relations are still
violated. This happens only marginally since for a slightly
positive value one could choose the free parameter, analogous to $x$
before, large enough to ensure that the uncertainty relation is
satisfied. It could happen that higher orders of the adiabatic
expansion would make the uncertainty product positive, but we do not
pursue this here since, if successful, it would most likely give only
special regimes where effective equations would be valid.

For $l=1$ we obtain
\[
 g_0^{cp}=\frac{\frac{5}{2}c^2-\frac{3}{2}pV(\phi)}{2(c^2-
 \frac{3}{4}pV(\phi))} g_0^{cc} \quad \mbox{and}\quad 
 g_0^{pp}= \frac{c^2}{c^2-\frac{3}{4}pV(\phi)} g_0^{cc}
\]
with a negative uncertainty product. For
$l=2$, we have
\[
 g_0^{cp}= \frac{c^2}{2(c^2-\frac{3}{4}pV(\phi))} g_0^{cc} 
 \quad \mbox{and} \quad
g_0^{pp}= \frac{c^4}{4(c^2-\frac{3}{4}pV(\phi))^2} g_0^{cc}
\]
with a vanishing uncertainty product. None of these cases gives a
consistent set of effective equations.

\subsubsection{Additive non-adiabaticity}

Instead of factorizing off the time dependence of the free solutions
as in the variables $g^{a,n}$, one can try to subtract it off, i.e.\
attempt an adiabatic expansion in $G^{cc}-C_3c^2$, $G^{cp}-C_4cp$ and
$G^{pp}-C_5p^2$. In contrast to the previous case the equations now
also depend on parameters $C_I$ of the free solutions. It is easy to
derive the resulting equations which, at any adiabatic order, provides
a set of inhomogeneous linear equations for the quantum
variables. There are three equations for three fluctuation parameters,
but with a vanishing determinant of the linear system. It turns out
that the inhomogeneity resulting from subtracting off the zeroth order
solutions does not allow any solution. Thus, also adiabaticity
additive to the zeroth order solutions is not realized.

\subsubsection{Non-adiabaticity with respect to free solutions}

Had one of the previous attempts succeeded in providing effective
equations, the result would have been time dependent only through the
scalar potential $V(\phi)$ as is clear from the examples provided. Our
final attempt brings in even stronger time dependence by assuming
again adiabaticity relative to free solutions, but using the free
solutions (\ref{FreeWDWI})--(\ref{FreeWDWV}) in explicit, time
dependent form $G_{V=0}^{a,n}(\phi)$ rather than implicitly as, e.g.,
$G^{cc}_{V=0}\propto c^2$ before. Thus, we now assume that variables
$g^{a,n}:=G^{a,n}/G_{V=0}^{a,n}(\phi)$ are adiabatic. The new
variables satisfy equations of motion
\[
 \dot{g}^{a,n}= \frac{\dot{G}^{a,n}}{G^{a,n}_{V=0}(\phi)}-
 \frac{\dot{G}^{a,n}_{V=0}(\phi)}{G^{a,n}_{V=0}(\phi)} g^{a,n}
\]
or explicitly
\begin{eqnarray}
 \dot{g}^{cc} &=& -2\left(1-
 \frac{c(c^2-pV(\phi)/2)}{(c^2-pV(\phi))^{3/2}}\right)g^{cc}- 2V(\phi)
\frac{c^2-3pV(\phi)/4}{(c^2-pV(\phi))^{3/2}}
\frac{G_{V=0}^{cp}(\phi)}{G_{V=0}^{cc}(\phi)} g^{cp}\,, \label{WDWTimeDepI}\\
  \dot{g}^{cc} &=& \frac{p^2V(\phi)}{(c^2-pV(\phi))^{3/2}}
\frac{G_{V=0}^{cc}(\phi)}{G_{V=0}^{cp}(\phi)} g^{cc}- V(\phi)
\frac{c^2-3pV(\phi)/4}{(c^2-pV(\phi))^{3/2}}
\frac{G_{V=0}^{pp}(\phi)}{G_{V=0}^{cp}(\phi)} g^{pp}\,,\label{WDWTimeDepII}\\
 \dot{g}^{pp} &=& 2\frac{p^2V(\phi)}{(c^2-pV(\phi))^{3/2}}
\frac{G_{V=0}^{cp}(\phi)}{G_{V=0}^{pp}(\phi)} g^{cp}+ 2\left(1-
\frac{c(c^2-pV(\phi)/2)}{(c^2-pV(\phi))^{3/2}}\right) g^{pp}\,.\label{WDWTimeDepIII}
\end{eqnarray}
To zeroth order, adiabatic solutions would be given by
\begin{eqnarray*}
 g_0^{cp} &=&
\frac{c(c^2-pV(\phi)/2)-(c^2-pV(\phi))^{3/2}}{V(\phi)(c^2-3pV(\phi)/4)}
\frac{G_{V=0}^{cc}(\phi)}{G_{V=0}^{cp}(\phi)} g_0^{cc}\,,\\
g_0^{pp} &=& \frac{p^2}{c^2-3pV(\phi)/4}
\frac{G_{V=0}^{cc}(\phi)}{G_{V=0}^{pp}(\phi)} g_0^{cc}
\end{eqnarray*}
which, when inserted into equations of motion for $c$ and $p$ would
give time dependent terms not only through $V(\phi)$ but also through
$G_{V=0}^{a,2}(\phi)$. This is an exponential time dependence and thus
stronger than that of a polynomial potential $V(\phi)$. However, also
here the result is not consistent since the uncertainty product
\[
 G_{V=0}^{cc}(\phi)g^{cc}G_{V=0}^{pp}(\phi)g^{pp}-(G_{V=0}^{cp}(\phi)g^{cp})^2\sim
-\frac{p^4V(\phi)^2}{64c^2(c^2-3pV(\phi)/4)^2}(G_{V=0}^{cc}(\phi))^2(g_0^{cc})^2
\]
to leading order in $pV(\phi)/c^2$ is again negative.

We did not manage to find a choice of adiabatic variables which would
give rise to consistent effective equations. However, since some cases
led only to marginal violations of the uncertainty relation, one may
still be hopeful that a choice exists. More importantly, changes to
the equations of motion such as those by a loop quantization could
possibly lead to the existence of consistent effective equations which
we will probe next.

\subsection{Effective equations of loop quantum cosmology}

The solvable loop model is formulated in non-canonical variables,
satisfying a Poisson algebra different from that of the
Wheeler--DeWitt model. Thus, also the procedure of deriving effective
equations is different and has to be analyzed anew.  We must now
consider two different sets of quantum variables, each one related to
$J$ or $\bar{J}$. We will first obtain general equations of motion for
these variables, which are then to be restricted by reality conditions
ensuring that $J\bar{J}=p^2$ or, at the quantum level,
$\hat{J}\hat{J}^{\dagger}=\hat{p}^2$. Expanded in powers of quantum
variables, the quantum Hamiltonian reads
\begin{eqnarray}\label{HQloop}
\nonumber H_Q &=& 
\frac{J-\bar{J}}{2i}-i\frac{p^3V(\phi)}{J-\bar{J}}-
i \frac{p^3V(\phi)}{(J-\bar{J})^3}  \left(
G^{JJ}+G^{\bar{J}\bar{J}}- 2G^{J\bar{J}} \right) \\ &&
+3i\frac{p^2V(\phi)}{(J-\bar{J})^2}  \left(
G^{pJ}-G^{p\bar{J}} \right) -3i \frac{pV(\phi)}{J-\bar{J}}
G^{pp} +\cdots \,.
\end{eqnarray}
From now on we ignore terms of higher moments since they will be
subdominant in a semiclassical expansion. Higher moments of typical
semiclassical states are smaller at higher orders. Moreover, a moment
of order $n$ appears in the quantum Hamiltonian with a coefficient
obtained by an $n$-th derivative of the classical Hamiltonian; see
(\ref{HQGen}). Higher moments of order $n$ are thus suppressed by
additional small factors of the form $p^{-k}(J-\bar{J})^{-n+k}$
compared to the classical perturbation term. There is a subtlety with
this argument when looking at equations of motion: In Poisson brackets
$\{G^{a,n},G^{b,m}\}$ with $n+m\geq 4$ (formally including $n=1$ or
$m=1$ as expectation values) there can not only be linear terms of
moments of order $n+m-1$, as in the basic relations
(\ref{pbracketII})--(\ref{JGJJbar}) below, but also terms of the form
$\langle\cdot\rangle^lG^{c,n+m-1-l}$ for $0\leq l\leq n+m-3$, as
illustrated by the examples in App.~\ref{App:Brack}. (Also
$\hbar^k$-terms appear, but they play a role only at higher orders.)
Depending on the powers $l$ realized, the expectation value dependent
prefactors could, in equations of motion, cancel suppressions involved
in coefficients of higher moment terms in the Hamiltonian. This is
potentially problematic if the highest allowed powers for $l$ do
occur. For instance, we have $\{G^{a,2},G^{b,2}\}$ of the form
$\langle\cdot\rangle G^{c,2}$ with a single factor of expectation
values, but in a bracket like $\{G^{a,2},G^{b,3}\}$ there could be
terms $\langle\cdot\rangle^2 G^{c,2}$ with up to two factors. Then,
the additional suppression of $G^{b,3}$ in the Hamiltonian would be
cancelled in equations of motion by the additional factor in Poisson
brackets. It would require one to take into account all terms
containing $G^{a,n}$ in the quantum Hamiltonian, which would clearly
be unmanageable.

Fortunately, this problem does not arise and there is a consistent way
to derive a truncation for equations of motion to any given order: For
equations of motion for expectation values, Poisson brackets
$\{\langle\cdot\rangle,G^{a,n}\}$ only have terms linear in moments
$G^{b,n}$ of the same order as used in the bracket, and possibly terms
$\langle\cdot\rangle G^{c,n-1}$ with a single factor of an expectation
value. (There may also be terms with higher powers of $\hbar$ which we
can safely ignore.) The terms with a single factor of expectation
values may arise because taking a commutator of, say, $\hat{p}$ with
$(\hat{J}-J)^n$ as it is needed for $\{p,G^{J^n}\}$ replaces one
factor of $\hat{J}-J$ by just $\hat{J}$ itself. To bring the result
into the form of a moment of order $n$, one has to add $J$ multiplied
with a moment of order $n-1$. The only other terms that may arise in
such cases are due to reordering, which contributes higher powers of
$\hbar$. (They would be relevant at higher moment orders than
considered here.) Taking into account suppression factors in the
Hamiltonian, equations of motion for expectation values thus have
suppressions of at least $\langle\cdot\rangle^{1-n}$ in higher moment
contributions. Being $n$-dependent, this leads to higher suppression
for higher moments and provides a consistent truncation. The same
order of truncation can then be extended to quantum variables by
requiring that the reality condition (\ref{reality}) is preserved by
the truncated equations of motion. This consistent truncation
procedure is illustrated below for moments of up to second order.

While all quantum variables commute with expectation values when
defined in terms of canonical variables, quantum variables and
expectation values do not Poisson commute for non-canonical variables
such as $p$ and $J$. Following the general scheme used before, Poisson
bracket relations between the variables are derived as
\begin{eqnarray}
 \{G^{J\bar{J}},p\} &=& 0 = \{G^{pp},p\} \label{pbracketII}\\
 \{G^{pJ},p\} &=& iG^{pJ} \quad,\quad \{G^{p\bar{J}},p\} =
-iG^{p\bar{J}} \label{pbracketI}\\
 \{G^{JJ},p\} &=& 2iG^{JJ}\quad,\quad \{G^{\bar{J}\bar{J}},p\} =
-2iG^{\bar{J}\bar{J}}
\end{eqnarray}
and
\begin{eqnarray}\label{brack-rel}
 \{ G^{pp}, J \} &=& -2iG^{pJ}\quad, \quad 
\{ G^{pp}, \bar{J} \} = 2i G^{p\bar{J}}\,, \label{JGpp}\\
\{G^{J\bar{J}},J\} &=& -2iG^{pJ}\quad,\quad \{G^{J\bar{J}},\bar{J}\} =
2iG^{p\bar{J}}\,, \label{JGJJ}\\
\{ G^{pJ}, J \} &=& -i G^{JJ}\quad, \quad 
\{ G^{pJ},\bar{J} \} = iG^{J\bar{J}}+ 2iG^{pp}\,, \\
\{ G^{p\bar{J}}, J \} &=& -iG^{J\bar{J}}-2iG^{pp}\quad,\quad
\{ G^{p\bar{J}}, \bar{J} \} = i G^{\bar{J}\bar{J}}\,,\\
\{ G^{JJ}, J \} &=& 0\quad, \quad \{G^{JJ}, \bar{J} \} = 4i G^{pJ}\,, \\
\{ G^{\bar{J}\bar{J}}, J \}&=& -4i G^{p\bar{J}}\quad, \quad 
\{ G^{\bar{J}\bar{J}}, \bar{J} \} = 0 \label{JGJJbar}
\end{eqnarray}
between quantum variables and expectation values. Notice that
(\ref{pbracketII}), (\ref{JGpp}) and (\ref{JGJJ}) imply that the
difference $G^{J\bar{J}}-G^{pp}$ commutes with any function of $p$,
$J$ and $\bar{J}$ only.  

Also in contrast to the Wheeler--DeWitt model whose canonical
variables gave rise to a closed Poisson algebra for the fluctuations
$G^{a,2}$, Poisson brackets between fluctuations in the loop model
involve higher moments due to the non-canonical form of the basic
variables $(p,J)$.  Poisson brackets between two fluctuations thus in
general depend on a quantum variable of order three, but they also
involve fluctuation dependent terms. While the third order variables
can be cut off in a semiclassical treatment to the order we are
interested in, the latter terms, as we will see below, have to be
included in a consistent set of effective equations. We thus need to
determine the full set of Poisson brackets between quantum variables
of order two. As derived in Appendix \ref{App:Brack}, they are
\begin{eqnarray}
 \{G^{pp},G^{J\bar{J}}\} &=& -2i(JG^{p\bar{J}}- \bar{J}G^{pJ})
\label{BracketppJbarJ}\\
 \{G^{pp},G^{JJ}\} &=& -4iG^{pJJ}-4iJG^{pJ}=
\overline{\{G^{pp},G^{\bar{J}\bar{J}}\}}\\
 \{G^{pp},G^{pJ}\} &=& -2iG^{ppJ}-2iJG^{pp}+{\textstyle\frac{1}{6}}i\hbar^2J=
\overline{\{G^{pp},G^{p\bar{J}}\}}\\
 \{G^{J\bar{J}},G^{JJ}\} &=& -4iG^{pJJ}-4i(p+\hbar/2)G^{JJ}=
\overline{\{G^{J\bar{J}},G^{\bar{J}\bar{J}}\}}\\
 \{G^{J\bar{J}},G^{pJ}\} &=& -2iG^{ppJ}\!-\!2i(p+\hbar/2)G^{pJ}
\!+\!iJG^{J\bar{J}}\!-\!i\bar{J}G^{JJ}\!+\! {\textstyle\frac{1}{6}}i\hbar^2J=
\overline{\{G^{J\bar{J}},G^{p\bar{J}}\}}\\
 \{G^{JJ},G^{pJ}\} &=& 2iG^{JJJ}+2iJG^{JJ}=
\overline{\{G^{\bar{J}\bar{J}},G^{p\bar{J}}\}}\\
 \{G^{JJ},G^{p\bar{J}}\} &=& 6iG^{ppJ}+8i(p+\hbar/2)G^{pJ}-
2i\bar{J}G^{JJ}-i\hbar^2J= \overline{\{G^{\bar{J}\bar{J}},G^{pJ}\}}\\
 \{G^{JJ},G^{\bar{J}\bar{J}}\} &=& 8iG^{ppp}+ 16i(p+\hbar/2)G^{pp}+
8i(p+\hbar/2)G^{J\bar{J}}- 8iJG^{p\bar{J}}-8i\bar{J}G^{pJ}\\
&&+4i\hbar^2p+2i\hbar^3\nonumber\\
 \{G^{pJ},G^{p\bar{J}}\} &=& 4iG^{ppp}+6i(p+\hbar/2)G^{pp}-
iJG^{p\bar{J}}-i\bar{J}G^{pJ}+{\textstyle\frac{1}{2}}i\hbar^2p+
{\textstyle\frac{1}{4}}i\hbar^3\,.\label{BracketpJpbarJ}
\end{eqnarray}
The relevant relations for leading order back-reaction effects in the
loop model are brackets of the form $\{G^{a,2},\langle\cdot\rangle\}$
together with contributions of $\{G^{a,2},G^{b,2}\}$ from moments up
to second order. (The fluctuation independent $\hbar^2$-terms can to
this order be ignored as well since $G^{a,2}$ are typically of order
$\hbar$ near saturation of the uncertainty relation.) Expectation
values and second moments thus provide a closed system under taking
Poisson brackets in this form, albeit a non-linear one.

\subsubsection{Equations of motion}

Just as the algebra (\ref{FluctPoissonI})--(\ref{FluctPoissonIII}) of
fluctuations in canonical variables is degenerate, being given by a
Poisson structure on a 3-dimensional manifold, the relations
$\{G^{a,2},\langle\cdot\rangle\}$ are degenerate in the sense that
functions of the fluctuations exist which commute with both
expectation values $p$ and $J$. It is easy to see that, as already
noted, one such function is linear, $G^{J\bar{J}}-G^{pp}=c_1$. The
second function is quadratic,
\begin{equation} \label{quadratic}
 C:=2(G^{pp})^2+(G^{J\bar{J}})^2-4|G^{p\bar{J}}|^2+ |G^{JJ}|^2\,.
\end{equation}
(Its value is $C=\frac{3}{4}c_1^2-\frac{1}{2}c_1c_2+\frac{3}{4}c_2^2
+4c_3c_4+4c_5c_6$ for the free solutions
(\ref{FreeLoopI})--(\ref{FreeLoopVI}).)  The linear function has one
immediate implication: It is exactly this combination which appears in
the reality condition (\ref{reality}) implied by
$\hat{J}\hat{J}^{\dagger}=\hat{p}^2$ in the loop quantization.  In the
free model, the combination $c_1=G^{J\bar{J}}-G^{pp}$ is a constant of
motion.  Whether or not this remains true when perturbations are
included is an interesting question because the form of the reality
condition (corresponding to an implementation of the correct physical
inner product) is essential for singularity resolution: it picks out
exactly the bouncing solutions in the free model (unless the constant
of motion $c_1$ would be negative and $|c_1|$ large, which is never
realized for states of the free model being semiclassical at least
once). If the form of the reality condition could be shown to be
preserved by quantum corrections, there awould be no obvious
violations of the bounce as one could in general imagine if $c_1$
would not remain constant and could evolve to large negative values
from initial semiclassical states \cite{BounceCohStates}. If, however,
$G^{J\bar{J}}-G^{pp}$ does not remain constant when the free model is
perturbed, one would have to be more careful with conclusions about
the bounce.

Since the difference $G^{J\bar{J}}-G^{pp}$ commutes with any function
of the expectation values, its preservation does not depend on the
precise form of the classical Hamiltonian. However, for leading
quantum corrections we have to consider fluctuation dependent terms in
the quantum Hamiltonian (\ref{HQloop}), which contribute extra terms
to the equations of motion of fluctuations. These fluctuation
dependent terms do not necessarily commute with $G^{J\bar{J}}-G^{pp}$.

Using the Poisson relations, we obtain
\begin{eqnarray}\label{eq-mot-loop}
\dot{p} &=& -\frac{J+\bar{J}}{2}+
\frac{J+\bar{J}}{(J-\bar{J})^2} p^3 V(\phi)+ 3
\frac{J+\bar{J}}{(J-\bar{J})^4}p^3
(G^{JJ}+G^{\bar{J}\bar{J}}-2G^{J\bar{J}})V(\phi)
 \\
\nonumber && -6 \frac{J+\bar{J}}{(J-\bar{J})^3}p^2 
(G^{pJ}- G^{p\bar{J}})V(\phi)+3 \frac{J+\bar{J}}{(J-\bar{J})^2}pG^{pp}\,V(\phi)
\\
&&- \frac{2p^3}{(J-\bar{J})^3} (G^{JJ}-G^{\bar{J}\bar{J}})V(\phi)+
\frac{3p^2}{(J-\bar{J})^2} (G^{pJ}+G^{p\bar{J}})V(\phi)\nonumber
\end{eqnarray}
and
\begin{eqnarray}
 \dot{J} &=& -(p+\hbar/2) + \left(\frac{3p^2J}{J-\bar{J}}+
\frac{2p^3(p+\hbar/2)}{(J-\bar{J})^2}\right)V(\phi)\label{eq-mot-loopJ}\\
&&+
\left(\frac{3p^2J}{(J-\bar{J})^3}+
\frac{6p^3(p+\hbar/2)}{(J-\bar{J})^4}\right)
(G^{JJ}+G^{\bar{J}\bar{J}}-2G^{J\bar{J}})V(\phi)\nonumber\\
&&-
3\left(\frac{2pJ}{(J-\bar{J})^2}+\frac{4p^2(p+\hbar/2)}{(J-\bar{J})^3}\right)
(G^{pJ}-G^{p\bar{J}})V(\phi)\nonumber\\
&&+ 3\left(\frac{J}{J-\bar{J}}+\frac{2p(p+\hbar/2)}{(J-\bar{J})^2}\right)
G^{pp}V(\phi)+ 4\frac{p^3}{(J-\bar{J})^3}(G^{p\bar{J}}-G^{pJ})V(\phi)
\nonumber\\
&&-
3\frac{p^2}{(J-\bar{J})^2} (G^{JJ}-G^{J\bar{J}}-2G^{pp})V(\phi)+
6\frac{p}{J-\bar{J}}G^{pJ}V(\phi) \nonumber
\end{eqnarray}
for the expectation values and
\begin{eqnarray} \label{EOMGpp}
\dot{G}^{pp} &=& -\left(1-\frac{2p^3}{(J-\bar{J})^2}\right)
(G^{pJ}+G^{p\bar{J}})V(\phi)\nonumber\\
&&- \frac{4p^3}{(J-\bar{J})^3}
\left(JG^{pJ}-\bar{J}G^{p\bar{J}}-
JG^{p\bar{J}}+\bar{J}G^{pJ}\right)V(\phi)\nonumber\\
&&+\frac{6p^2(J+\bar{J})}{(J-\bar{J})^2}G^{pp}V(\phi)-
\frac{\hbar^2}{2}\frac{p^2(J+\bar{J})}{(J-\bar{J})^2}V(\phi)\nonumber\\
&=&-\left(1-\frac{2p^3}{(J-\bar{J})^2}\right)
(G^{pJ}+G^{p\bar{J}})V(\phi)- \frac{4p^3(J+\bar{J})}{(J-\bar{J})^3}
(G^{pJ}-G^{p\bar{J}})V(\phi)\nonumber\\
&&+\frac{6p^2(J+\bar{J})}{(J-\bar{J})^2}G^{pp}V(\phi)-
\frac{\hbar^2}{2}\frac{p^2(J+\bar{J})}{(J-\bar{J})^2}V(\phi)
\end{eqnarray}
and
\begin{eqnarray} \label{EOMGJbarJ}
 \dot{G}^{J\bar{J}} &=& -\left(1-\frac{2p^3}{(J-\bar{J})^2}\right)
(G^{pJ}+G^{p\bar{J}})V(\phi)- \frac{4p^3(p+\hbar/2)}{(J-\bar{J})^3}
(G^{JJ}-G^{\bar{J}\bar{J}})V(\phi)\nonumber\\
 &&- \frac{3p^2}{(J-\bar{J})^2}
((J+\bar{J})G^{J\bar{J}}- (2p+\hbar)(G^{pJ}+G^{p\bar{J}})-
\bar{J}G^{JJ}-JG^{\bar{J}\bar{J}}\nonumber\\
&&+{\textstyle\frac{1}{6}}\hbar^2(J+\bar{J})) V(\phi)
+\frac{6p}{J-\bar{J}}(JG^{p\bar{J}}-\bar{J}G^{pJ})V(\phi)\nonumber\\
&=& -\left(1-3p-\frac{2p^2(p+3(p+\hbar/2))}{(J-\bar{J})^2}\right)
 (G^{pJ}+G^{p\bar{J}})V(\phi)\nonumber\\
&& -\frac{3p(J+\bar{J})}{J-\bar{J}} 
(G^{pJ}-G^{p\bar{J}})V(\phi)
 - \frac{3}{2} \frac{p^2}{J-\bar{J}}
\left(1+\frac{8}{3}\frac{p(p+\hbar/2)}{(J-\bar{J})^2}\right) 
(G^{JJ}-G^{\bar{J}\bar{J}})V(\phi)\nonumber\\
&& +\frac{3}{2} 
\frac{p^2(J+\bar{J})}{(J-\bar{J})^2}
(G^{JJ}+G^{\bar{J}\bar{J}}- 2G^{J\bar{J}})V(\phi)
-\frac{\hbar^2}{2} \frac{p^2(J+\bar{J})}{(J-\bar{J})^2}V(\phi)
\end{eqnarray}
for the fluctuations featuring in the reality condition, ignoring
terms quadratic in fluctuations. A combination gives
\begin{eqnarray}
 \dot{G}^{J\bar{J}}-\dot{G}^{pp} &=& p
\left(3+
\frac{6p(p+\hbar/2)}{(J-\bar{J})^2}\right)
(G^{pJ}+G^{p\bar{J}})V(\phi)\\
&&+ \frac{p(J+\bar{J})}{J-\bar{J}}
\left(-3+4\frac{p^2}{(J-\bar{J})^2}\right) (G^{pJ}-G^{p\bar{J}})V(\phi)\nonumber\\
&&-\frac{p^2}{J-\bar{J}}\left(\frac{3}{2}+
\frac{4p(p+\hbar/2)}{(J-\bar{J})^2}\right) (G^{JJ}-G^{\bar{J}\bar{J}})V(\phi)
\nonumber\\
&&+\frac{3}{2} \frac{p^2(J+\bar{J})}{(J-\bar{J})^2}
(G^{JJ}+G^{\bar{J}\bar{J}}- 2G^{J\bar{J}}-4G^{pp})V(\phi)\nonumber
\end{eqnarray}
which can be seen to equal $2\md (|J|^2-(p+\hbar/2)^2)/\md \phi$ from
(\ref{eq-mot-loop}) and (\ref{eq-mot-loopJ}). Terms included in these
equations of motion preserve the reality condition. Thus, this
consistent truncation of the quantum equations describes the physical
evolution of expectation values and fluctuations as they would be
computed from a physical state.

However, these equations also show that the contribution
$G^{J\bar{J}}-G^{pp}$ to the reality condition is no longer constant
when fluctuation dependent corrections in $H_Q$ are allowed for.  It
is then possible that the term $c_1$, which was constant in the free
model and determines whether or not expectation values bounce, changes
in time so as to prevent a bounce when small volume is approached during
long evolution. A firm conclusion about the presence of a bounce for a
massive or interacting scalar thus requires a more detailed analysis
taking into account the behavior of fluctuations. We will come back to
this issue in the next section.

Equations of motion for the remaining second moments are
\begin{eqnarray} \label{EOMGpJ}
\dot{G}^{pJ} &=& -\frac{1}{2}\left(1-\frac{2p^3}{(J-\bar{J})^2}\right)
(G^{JJ}+G^{J\bar{J}}+2G^{pp})V(\phi)
+\frac{3p^2}{J-\bar{J}}G^{pJ}V(\phi)\\
&&+\frac{p^3V(\phi)}{(J-\bar{J})^3} (-2(J+\bar{J})G^{JJ}-2JG^{\bar{J}\bar{J}}
+2(p+\hbar/2)(4G^{p\bar{J}}-2G^{pJ}) +2JG^{J\bar{J}}\nonumber\\
&&
-\hbar^2\bar{J}+{\textstyle \frac{1}{3}}\hbar^2J)
+\frac{3p^2V(\phi)}{(J-\bar{J})^2}
(6(p+\hbar/2)G^{pp}-\bar{J}G^{pJ}-JG^{p\bar{J}}
+{\textstyle\frac{1}{2}}\hbar^2p+{\textstyle\frac{1}{4}}\hbar^3)\nonumber\\
&&+
\frac{3pV(\phi)}{J-\bar{J}} (2JG^{pp}-{\textstyle\frac{1}{6}}\hbar^2J)\nonumber\\
&=& \overline{\dot{G}^{p\bar{J}}} \nonumber
\end{eqnarray}
and
\begin{eqnarray}\label{EOMGJJ}
\dot{G}^{JJ} &=& -2\left(1-\frac{2p^3}{(J-\bar{J})^2}\right)G^{pJ}V(\phi)
+\frac{6p^2}{J-\bar{J}}G^{JJ}V(\phi)\\
&&+ \frac{p^3V(\phi)}{(J-\bar{J})^3}
(8(p+\hbar/2)(2G^{pp}+ G^{J\bar{J}}-G^{JJ})- 8JG^{p\bar{J}}
-8\bar{J}G^{pJ} +4\hbar^2p+2\hbar^3) \nonumber\\
&&-\frac{3p^2V(\phi)}{(J-\bar{J})^2} (2JG^{JJ}- 8(p+\hbar/2)G^{pJ}
+2\bar{J}G^{JJ}+\hbar^2J)
+\frac{12p}{J-\bar{J}}JG^{pJ}V(\phi)\nonumber\\
 &=& \overline{\dot{G}^{\bar{J}\bar{J}}}\,. \nonumber
\end{eqnarray}
As in the Wheeler--DeWitt context, we will now briefly describe
attempts to find adiabatic regimes of quantum variables, followed in
the next section by a perturbative analysis of the equations of
motion.

\subsubsection{Non-adiabaticity of quantum variables}

For simplicity, we now take into account the classical Hamiltonian as
a function of expectation values only in order to compute the leading
order terms in equations of motion for fluctuations. These equations
are
\begin{eqnarray}
 \dot{G}^{pp} &=& -\left(1- \frac{2p^3}{(J-\bar{J})^2}V(\phi)\right)
(G^{pJ}+G^{p\bar{J}})= \dot{G}^{J\bar{J}}\,, \label{AdLoopQuantI}\\
 \dot{G}^{pJ} &=& -\frac{1}{2}\left(1-
\frac{2p^3}{(J-\bar{J})^2}V(\phi)\right) (G^{JJ}+G^{J\bar{J}}+2G^{pp})
+\frac{3p^2}{J-\bar{J}}V(\phi)G^{pJ}=
\overline{\dot{G}^{p\bar{J}}}\,,\label{AdLoopQuantII}\\
\dot{G}^{JJ} &=& -2\left(1-\frac{2p^3}{(J-\bar{J})^2}V(\phi)\right)
G^{pJ}+ \frac{6p^2}{J-\bar{J}}V(\phi)G^{JJ}=
\overline{\dot{G}^{\bar{J}\bar{J}}}\,.\label{AdLoopQuantIII}
\end{eqnarray}
These are not the complete equations of motion; we use them only to
illustrate obstacles to the existence of adiabatic effective
equations. The full equations of motion to this order will be used
later on for a new type of effective system.

To zeroth order in an adiabatic expansion of the fluctuations,
Eq.~(\ref{AdLoopQuantI}) shows that $G_0^{pJ}$ is purely imaginary,
from which Eq.~(\ref{AdLoopQuantIII}) then shows that $G_0^{JJ}$ is
real. Moreover, Eq.~(\ref{AdLoopQuantIII}) gives the relation between
these two quantities,
\begin{equation} \label{GpJZero}
 G_0^{pJ}-G_0^{p\bar{J}}=
\frac{3p^2V(\phi)}{1-2p^3V(\phi)/(J-\bar{J})^2}
\frac{G_0^{JJ}+G_0^{\bar{J}\bar{J}}}{J-\bar{J}}
\end{equation}
which when used in Eq.~(\ref{AdLoopQuantII}) leaves only one free
function such as $G_0^{pp}$: to leading order in
$p^2V(\phi)/(J-\bar{J})$ we have
\begin{equation} \label{Gcomb}
 G_0^{JJ}+G_0^{\bar{J}\bar{J}}+6G_0^{pp}+2c_1=
\frac{18p^4V(\phi)^2}{(J-\bar{J})^2} (G_0^{JJ}+G_0^{\bar{J}\bar{J}})
\end{equation}
and thus
\begin{equation} \label{GJJzero}
 G_0^{JJ}+G_0^{\bar{J}\bar{J}}=
-\frac{6G_0^{pp}+2c_1}{1-18p^4V(\phi)^2/(J-\bar{J})^2}\,.
\end{equation}
In these equations, also $G_0^{J\bar{J}}-G_0^{pp}=c_1$ has been
used. Unlike in the Wheeler--DeWitt case, the remaining free function
$G_0^{pp}$ would not be fixed at the next order in the adiabatic
expansion but through the second, quadratic constant $C$ in
(\ref{quadratic}).

We first verify that the fluctuations are at least positive as they
should by definition.  While this does not apply to $G^{JJ}$,
$\hat{J}$ not being self-adjoint, the quantity
$G^{J+\bar{J},J+\bar{J}}:=\langle(\hat{J}+\hat{J}^{\dagger})^2\rangle-
\langle\hat{J}+\hat{J}^{\dagger}\rangle^2$ must satisfy positivity
conditions. Eq.~(\ref{GJJzero}) in fact shows that
$G_0^{JJ}+G_0^{\bar{J}\bar{J}}$ is negative, and thus
$G_0^{JJ}+G_0^{\bar{J}\bar{J}}+6G_0^{pp}+2c_1$ is positive from
(\ref{Gcomb}). This is relevant because the latter quantity can be
written as $G^{JJ}+G^{\bar{J}\bar{J}}+6G^{pp}+2c_1=
G^{J+\bar{J},J+\bar{J}}+4G^{pp}$ in terms of the positive fluctuation
$G^{J+\bar{J},J+\bar{J}}:=\langle(\hat{J}+\hat{J}^{\dagger})^2\rangle-
\langle\hat{J}+\hat{J}^{\dagger}\rangle^2$ of the self-adjoint
$\hat{J}+\hat{J}^{\dagger}$. The zeroth order adiabatic solutions are
thus not in conflict with positivity of fluctuations.

Next, we check uncertainty relations. This time, there are several
independent ones \cite{BounceCohStates} because we are using partially
complex basic variables. Independent uncertainty relations must then
be fulfilled for each pair $(\hat{p},\hat{J}+\hat{J}^{\dagger})$,
$(\hat{p},i(\hat{J}-\hat{J}^{\dagger})$ and
$\hat{J}+\hat{J}^{\dagger}, i(\hat{J}-\hat{J}^{\dagger})$ of
self-adjoint operators. These relations, derived in
\cite{BounceCohStates}, take the form
\begin{eqnarray}
 G^{pp}G^{J+\bar{J},J+\bar{J}}- (G^{p,J+\bar{J}})^2
  &\geq& -\frac{1}{4}\hbar^2(J-\bar{J})^2\label{uncertI}\\
 G^{pp}G^{i(J-\bar{J}),i(J-\bar{J})}- (G^{p,i(J-\bar{J})})^2
 &\geq& \frac{1}{4}\hbar^2(J+\bar{J})^2\label{uncertII}\\
 G^{J+\bar{J},J+\bar{J}}G^{i(J-\bar{J}),i(J-\bar{J})}-
 (G^{J+\bar{J},i(J-\bar{J})})^2
 &\geq& \hbar^2(2p+\hbar)^2\,.\label{uncertIII}
\end{eqnarray}
The first one is easy to satisfy to this order since $G^{p,J+\bar{J}}=
G^{pJ}+G^{p\bar{J}}$ vanishes. The left hand side of (\ref{uncertI})
is then positive as just shown, and one can find appropriate scalings
(i.e.\ sufficiently large $C$) to ensure the uncertainty relation
noting that $J-\bar{J}$ is nearly constant for perturbative solutions.
The second relation is more difficult to verify since
$G^{p,i(J-\bar{J})}$ is non-zero. For the zeroth order solutions, we
have
\[
 G_0^{pp}G_0^{i(J-\bar{J}),i(J-\bar{J})}- (G_0^{p,i(J-\bar{J})})^2=
 -G_0^{pp}(G_0^{JJ}+G_0^{\bar{J}\bar{J}}-2 G_0^{J\bar{J}})+
 (G_0^{pJ}-G_0^{p\bar{J}})^2
\]
which is at least positive since we saw that
$G_0^{JJ}+G_0^{\bar{J}\bar{J}}$ is negative and
$(G_0^{pJ}-G_0^{p\bar{J}})^2$, although negative, is of order
$V(\phi)^2$ as a consequence of (\ref{GpJZero}). Moreover, using the
free solutions as a guide indicates that one might be able to satisfy
the uncertainty relation: In the free case, we have
\[
 -(G^{JJ}_{V=0}+G^{J\bar{J}}_{V=0}-2 G^{J\bar{J}}_{V=0})= 2(c_1-c_2)=
 4(\Delta H)^2
\]
which is positive and constant. With the factor of $G^{pp}$, which in
the free case behaves as $p^2$ at least far away from the bounce one
may have a chance to satisfy (\ref{uncertII}). Similarly, the third
relation, which is simpler due to $G_0^{J+\bar{J},i(J-\bar{J})}=
i(G_0^{JJ}-G_0^{\bar{J}\bar{J}})=0$, has a left hand side
\begin{eqnarray*}
 G_0^{J+\bar{J},J+\bar{J}}G_0^{i(J-\bar{J}),i(J-\bar{J})}-
 (G_0^{J+\bar{J},i(J-\bar{J})})^2 &=& -(G_0^{JJ}+G_0^{\bar{J}\bar{J}})^2+
 4(G_0^{J\bar{J}})^2\\
 &=& -\frac{(6G_0^{pp}+2c_1)^2}{(1-18p^4V(\phi)^2/(J-\bar{J})^2)^2}+
 4(G^{pp}_0+c_1)^2\\
 &=& -32 (G^{pp}_0)^2 - 16c_1 G_0^{pp}- 144
\frac{p^4V(\phi)^2}{(J-\bar{J})^2} (3G_0^{pp}+c_1)^2
\end{eqnarray*}
for the first two leading orders in $p^2V(\phi)/(J-\bar{J})$. It can
be positive for states satisfying $c_1<-2G_0^{pp}$, but also has to be
larger than the $\phi$-dependent $4\hbar^2p^2$ as required by the
uncertainty relation. A look at the free solutions also here indicates
that this might be possible: both $c_1$ and $G^{pp}_{V=0}$ are of the
order $\hbar$  for saturation of the free uncertainty relations, and
$G^{pp}_{V=0}$ behaves as $p^2$ far away from the bounce.

Using the candidate solutions for adiabatic quantum variables in the
equation of motion for $p$ then suggests an effective equation
\begin{eqnarray}
 \dot{p} &=& -\frac{J+\bar{J}}{2}
+\frac{p^3(J+\bar{J})}{(J-\bar{J})^2} V(\phi)
+3\frac{J+\bar{J}}{(J-\bar{J})^2}
\left(1-8\frac{p^3}{(J-\bar{J})^2}\right) G_0^{pp}V(\phi)\label{peffLoop}\\
&&-
12\frac{p^3(J+\bar{J})}{(J-\bar{J})^4} c_1 V(\phi)
+36\frac{p^4(J+\bar{J})}{(J-\bar{J})^4} (3G_0^{pp}+c_1) V(\phi)^2\nonumber
\end{eqnarray}
up to higher order terms in the potential. Here, $c_1$ is a constant
parameterizing the state around which the effective equation is
derived. Similarly, $G_0^{pp}$ depends on $c_1$ and the second
constant $C$ in (\ref{quadratic}) which is preserved by the equations
of motion we consider here. Thus, we would have an effective equation
containing only two state dependent parameters but no free
functions. However, in the final step of determining $G^{pp}_0(c_1,C)$
an inconsistency arises: using the candidate adiabatic solutions in
(\ref{quadratic}) for a constant $C$, $G^{pp}_0$ has to be constant up
to perturbative terms of the order $p^4V^2/(J-\bar{J})^2$. However, we
saw that the uncertainty relations require $G^{pp}_0$ to behave as
$p^2$ as it does in the free case. With a potential, there is no
adiabatic regime in which this can be satisfied and the effective
equation (\ref{peffLoop}) is, after all, inconsistent.

\subsubsection{Non-adiabaticity relative to free solutions}

For the loop quantized model it is more complicated to introduce
adiabaticity relative to free solutions because free quantum variables
(\ref{FreeLoopI})--(\ref{FreeLoopVI}) are not simply proportional to
powers of the expectation values as in the Wheeler--DeWitt case. The
most direct possibility is to introduce the free solutions as
explicitly time dependent functions $G^{a,n}_{V=0}(\phi)$. Resulting
effective equations will then be strongly time dependent not just
through the scalar potential. Then, for variables $g^{a,n}:=
G^{a,n}/G^{a,n}_{V=0}(\phi)$ and using
(\ref{AdLoopQuantI})--(\ref{AdLoopQuantIII}) we have equations of
motion
\begin{eqnarray*}
 \dot{g}^{pp} &=& -\left(1-\frac{2p^3V(\phi)}{(J-\bar{J})^2}\right)
\left(\frac{G^{pJ}_{V=0}(\phi)}{G^{pp}_{V=0}(\phi)}
g^{pJ}+\frac{G^{p\bar{J}}_{V=0}(\phi)}{G^{pp}_{V=0}(\phi)} g^{p\bar{J}}\right)+
\frac{G^{pJ}_{V=0}(\phi)+G^{p\bar{J}}_{V=0}(\phi)}{G^{pp}_{V=0}(\phi)} g^{pp}\,,\\
 \dot{g}^{pJ} &=& -\frac{1}{2}
\left(1-\frac{2p^3V(\phi)}{(J-\bar{J})^2}\right)
\left(\frac{G^{JJ}_{V=0}(\phi)}{G^{pJ}_{V=0}(\phi)}
g^{JJ}+\frac{G^{J\bar{J}}_{V=0}(\phi)}{G^{pJ}_{V=0}(\phi)}
g^{J\bar{J}}+2\frac{G^{pp}_{V=0}(\phi)}{G^{pJ}_{V=0}(\phi)}
g^{pp}\right)\nonumber\\
&&+ \frac{1}{2}
\left(\frac{G^{JJ}_{V=0}+G^{J\bar{J}}_{V=0}(\phi)+2G^{pp}_{V=0}(\phi)}{G^{pJ}_{V=0}(\phi)}+
\frac{6p^2V(\phi)}{J-\bar{J}}\right) g^{pJ}=
\overline{\dot{g}^{p\bar{J}}}\,,\\
 \dot{g}^{JJ} &=& -2 \left(1-\frac{2p^3V(\phi)}{(J-\bar{J})^2}\right)
\frac{G^{pJ}_{V=0}(\phi)}{G^{JJ}_{V=0}(\phi)} g^{pJ}+
2\left(\frac{G^{pJ}_{V=0}(\phi)}{G^{JJ}_{V=0}(\phi)}+
\frac{3p^2V(\phi)}{J-\bar{J}}\right) g^{JJ}=
\overline{\dot{g}^{\bar{J}\bar{J}}}\,,\\
\dot{g}^{J\bar{J}} &=&
-\left(1-\frac{2p^3V(\phi)}{(J-\bar{J})^2}\right)
\left(\frac{G^{pJ}_{V=0}(\phi)}{G^{J\bar{J}}_{V=0(\phi)}} g^{pJ}+
\frac{G^{p\bar{J}}_{V=0}(\phi)}{G^{J\bar{J}}_{V=0}(\phi)} g^{p\bar{J}}\right)+
\frac{G^{pJ}_{V=0}(\phi)+G^{p\bar{J}}_{V=0}(\phi)}{G^{J\bar{J}}_{V=0}(\phi)}
g^{J\bar{J}}\,.
\end{eqnarray*}

There is now a crucial difference to the situation in a
Wheeler--DeWitt quantization as in Eqs.~(\ref{WDWTimeDepI}),
(\ref{WDWTimeDepII}) and (\ref{WDWTimeDepIII}): $V=0$ is not a
singular point for the adiabatic approximation. In the Wheeler--DeWitt
equations, some terms completely vanish in the free case such that the
solutions change dramatically. For instance, solving for $g_0^{cp}$
requires one to divide by $V(\phi)$ which is not possible in the free
case. Thus, the existence of consistent solutions satisfying
uncertainty relations in the free case, and obviously being adiabatic
for the variables $g^{a,n}=G^{a,n}/G^{a,n}_{V=0}(\phi)=1$, does not
guarantee the existence of consistent adiabatic solutions even under
slight perturbations including a small potential. The equations
obtained here for the loop quantization, on the other hand, are
perfectly perturbative in $V(\phi)$ and the free solutions do not
present a singular point for the adiabatic expansion.

Nevertheless, there are again no adiabatic solutions in the presence
of a potential. While the linear system of six equations for six
(real) variables at zeroth adiabatic order is degenerate when $V=0$,
allowing a non-trivial solution where all $g^{a,n}=1$, the determinant
of the coefficients can be seen to be non-vanishing for
$V\not=0$. Thus, the only solution in this case is $g^{a,n}=0$ which
again violates the uncertainty principles.

There might be adiabatic regimes for the equations
(\ref{eq-mot-loop})--(\ref{EOMGJbarJ}) and
(\ref{EOMGpJ})--(\ref{EOMGJJ}) to this order, but this would be more
involved and does not seem promising. We therefore turn to a new type
of effective equations which does not require an adiabatic expansion.

\section{Higher dimensional effective equations and perturbative solutions}

Although our attempts to find adiabatic regimes and to solve for
quantum variables in terms of expectation values were largely
unsuccessful, prohibiting the derivation of effective equations solely
in terms of expectation values, effective equations for an enlarged
system do exist. Rather than solving for the fluctuations, we keep
them as independent variables in addition to the expectation
values. This is still a system of classical type, given in terms of
finitely many variables, since we already removed higher moments as
part of the semiclassical expansion. But it is certainly different
from the exact classical system not only in the form of its equations
of motion but also by the number of independent variables. From a
general viewpoint, this is in agreement with expectations from
effective actions which generically give rise to higher derivative
terms. Higher derivative actions imply higher derivative equations of
motion which, when taken at face value, describe a larger set of
independent degrees of freedom: additional parameters have to be
specified for a complete initial value problem of higher derivative
equations. However, when higher derivative terms arise from a
perturbation expansion not all the solutions are consistent within
this scheme. Only solutions analytic in the perturbation parameter are
consistent, which can be shown to allow a number of independent
solutions agreeing with the unperturbed case
\cite{SingHighOrder,Simon}. Thus, higher derivative effective actions
do not truly introduce new degrees of freedom. Keeping some of the
quantum variables, on the other hand, requires additional initial
conditions even for perturbative solutions; in this case new quantum
degrees of freedom truly arise.

\subsection{Wheeler--DeWitt model}

In variables $c$ and $p$, we thus have a set of five coupled effective
equations
\begin{eqnarray}\label{eq-mot-class}
\nonumber \dot{c}&=& c+\left(- \frac{p}{c}-\frac{p}{c^3} G^{cc}+ 
\frac{1}{c^2}  G^{cp} \right)V(\phi) \\
\nonumber \dot{p}&=& -p+\left( -\frac{p^2}{2c^2}-\frac{3p^2}{2c^4}
G^{cc}+ \frac{2p}{c^3} G^{cp} -\frac{1}{2c^2}  G^{pp} \right) V(\phi)\\
\nonumber \dot{G}^{cc}&=& 2 \left(1+ \frac{p}{c^2}V(\phi) \right)G^{cc}- 
\frac{2}{c} G^{cp} V(\phi)\\
\nonumber \dot{G}^{cp}&=& \left( \frac{p^2}{c^3}G^{cc}- 
\frac{1}{c}G^{pp} \right) V(\phi)\\
\dot{G}^{pp} &=& \frac{2p^2}{c^3} G^{cp} V(\phi)- 
2\left(1+\frac{p}{c^2}V(\phi) \right) G^{pp}.
\end{eqnarray}
Back-reaction terms of fluctuations are included in the
equations of motion for expectation values, although the role of an
effective potential is illustrated more indirectly than it would be
for adiabatic effective equations.

For perturbations around free solutions, assuming small values of the
potential $V(\phi)=\epsilon {\cal V}(\phi)$ with a small parameter
$\epsilon$, we define
\begin{equation} \label{pexpand}
p(\phi)= p_0(\phi)+\epsilon p_1(\phi)+ \cdots
\end{equation}
and similarly for the other variables, where $p_0$ is the zeroth order
solution (\ref{FreeWDWII}) corresponding to $V(\phi)=0$. (For
simplicity we now use a simple zero as subscript for the free
solutions rather than ``$V=0$''. Confusion will not arise because the
adiabatic approximation will no longer be used.) All free solutions
are explicitly given by (\ref{FreeWDWI})--(\ref{FreeWDWV}). To first
perturbative order in $\epsilon$, the equations of motion
(\ref{eq-mot-class}) become
\begin{eqnarray}\label{eq-mot-pert-class}
\nonumber \dot{c}_1 &=& c_1+ \left(-\frac{p_0}{c_0}-
\frac{p_0}{c_0^3}G_0^{cc}+\frac{1}{c_0^2} G_0^{cp} \right) {\cal V}(\phi) \\
\nonumber \dot{p}_1 &=& -p_1+ \left( -\frac{1}{2} 
\frac{p_0^2}{c_0^2}- \frac{3p_0^2}{2c_0^4}G_0^{cc}+2 
\frac{p_0}{c_0^3} G_0^{cp}- \frac{1}{2c_0^2}G_0^{pp} \right) {\cal V}(\phi) \\
\nonumber \dot{G}_1^{cc} &=& 2 G_1^{cc}+ 2\left( \frac{p_0}
{c_0^2} G_0^{cc}- \frac{1}{c_0} G_0^{cp} \right) {\cal V}(\phi) \\
\nonumber \dot{G}_1^{cp} &=& \left( \frac{p_0^2}
{c_0^3} G_0^{cc} -\frac{1}{c_0} G_0^{pp} \right) {\cal V}(\phi) \\
\dot{G}_1^{pp} &=& -2 G_1^{pp}+ 2\left( \frac{ p_0^2}{c_0^3} 
G_0^{cp}-\frac{ p_0}{c_0^2} G_0^{pp} \right) {\cal V}(\phi)
\end{eqnarray}
which are inhomogeneous differential equations whose homogeneous terms
are linear. Proceeding in the same way to higher orders, this provides
a well-defined and self-consistent solution scheme for the interacting
dynamics of the quantum system. There are two expansions of the full
quantum equations: a semiclassical one (truncating higher moments) and
the perturbative one in $\epsilon$ (removing higher powers of the
potential). When extending to higher orders, one would thus have to
find out what the relative magnitudes of correction terms in the
different expansions are.

Using the free solutions (\ref{FreeWDWI})--(\ref{FreeWDWV}) for the
different variables we obtain the perturbative solutions up to first
order in $\epsilon$:
\begin{eqnarray*}
 c_1(\phi) &=& B_1 e^{\phi}+
(C_2/C_1+C_2C_3/C_1^3-C_4/C_1^2) e^{\phi}\int^{\phi} {\cal
V}(\tau)e^{-3\tau}\md\tau\\
p_1(\phi) &=& B_2 e^{-\phi}- (-C_2^2/2C_1^2-3C_2^2C_3/2C_1^4+
2C_2C_4/C_1^3-C_5/2C_1^2)e^{-\phi} \int^{\phi} {\cal
V}(\tau)e^{-3\tau}\md\tau\\
G_1^{cc}(\phi) &=& B_3 e^{2\phi} - 2(C_2C_3/C_1^2- C_4/C_1)
e^{2\phi} \int^{\phi} {\cal V}(\tau)e^{-3\tau}\md\tau\\
G_1^{cp}(\phi) &=& B_4- (C_2^2C_3/C_1^3-C_5/C_1) \int^{\phi} {\cal
V}(\tau)e^{-3\tau}\md\tau\\
G_1^{pp}(\phi) &=& B_5 e^{-2\phi} - 2(C_2^2C_4/C_1^3-C_2C_5/C_1^2)
e^{-2\phi} \int^{\phi} {\cal V}(\tau)e^{-3\tau}\md\tau
\end{eqnarray*}
where $(B_1, \ldots, B_5)$ are further constants of integration simply
adding up to the zeroth order constants in
(\ref{FreeWDWI})--(\ref{FreeWDWV}).  Specifically, for a mass term
potential $V(\phi)=\frac{1}{2}m^2\phi^2=\epsilon \phi^2$ we have
\begin{eqnarray*}
c_1(\phi) &=& B_1 e^{\phi}+
\frac{1}{27}(C_2/C_1+C_2C_3/C_1^3-C_4/C_1^2) (2+6\phi+9\phi^2)e^{-2\phi}\\
p_1(\phi) &=& B_2 e^{-\phi}- \frac{1}{27} (-C_2^2/2C_1^2-3C_2^2C_3/2C_1^4+
2C_2C_4/C_1^3-C_5/2C_1^2) (2+6\phi+9\phi^2)e^{-4\phi}\\
G_1^{cc}(\phi) &=& B_3 e^{2\phi} -\frac{2}{27} (C_2C_3/C_1^2- C_4/C_1)
(2+6\phi+9\phi^2)e^{-\phi} \\
G_1^{cp}(\phi) &=& B_4- \frac{1}{27} (C_2^2C_3/C_1^3-C_5/C_1)
(2+6\phi+9\phi^2)e^{-3\phi}\\
G_1^{pp}(\phi) &=& B_5 e^{-2\phi} - 
\frac{2}{27} (C_2^2C_4/C_1^3-C_2C_5/C_1^2)
(2+6\phi+9\phi^2)e^{-5\phi}\,.
\end{eqnarray*}

These solutions already display interesting properties of the
system. Compared to the free solutions which are pure exponentials in
$\phi$, there are different types of corrections. Since we are
perturbatively adding a potential one certainly expects corrections to
the solutions which correspond to some of the polynomial terms. For
the classical variables $c$ and $p$, these agree with corrections to
the classical solutions resulting from the potential. But there are
additional corrections displaying genuine quantum behavior since
fluctuations $G^{a,2}$ evolve, too, and couple to the expectation
values. Although they are structurally similar, one can easily
disentangle both types of corrections by setting the zeroth order
fluctuations to zero. (This is not possible for the quantum system
since it would violate uncertainty relations. But we are free to do
so formally to discuss different correction terms.)  Zeroth order
fluctuations are zero if we set $C_3=C_4=C_5=0$.  This removes some
corrections from $c_1$ and $p_1$ which can therefore be attributed
unambiguously to quantum back-reaction. They differ from classical
corrections only in their coefficients, related to the integration
constants. The time dependence of the correction, in this model, is
thus the same as the classical one and quantum back-reaction does not
change the expectation values significantly. Moreover, their
coefficients are small if a state is semiclassical at large volume,
which requires that the fluctuation parameters $C_3$, $C_4$ and $C_5$
are small compared to $C_1$ and $C_2$.

\subsection{Loop quantization}

The loop quantization, being formulated in non-canonical variables
$(p,J)$, leads to a different algebra of expectation values and
quantum variables. Moreover, due to the use of the complex variable
$J$, which is typical for loop quantization based on holonomies, there
are more independent variables and equations of motion. This is
supplemented by a reality condition (\ref{reality}) which leaves the
correct number of degrees of freedom when implemented. The Poisson
brackets (\ref{pbracketI})--(\ref{JGJJbar}) together with the quantum
Hamiltonian (\ref{HQloop}) result in equations of motion
(\ref{eq-mot-loop}) and (\ref{eq-mot-loopJ}).  With equations of
motion (\ref{EOMGpp}), (\ref{EOMGJbarJ}), (\ref{EOMGpJ}) and
(\ref{EOMGJJ}) for $\dot{G}^{a,2}$, this provides our higher dimensional
effective system of the loop quantization.

For perturbative solutions of the expectation values we expand as in
(\ref{pexpand}), use free solutions $p_0(\phi)$, $J_0(\phi)$ and
$G^{a,2}_0(\phi)$ from (\ref{FreeWDWI})--(\ref{FreeWDWV}) on the right
hand side and integrate. We do not require solutions or even equations
of motion for $G_1^{a,2}$ to this order in a perturbative solution
scheme if we are only interested in $p_1(\phi)$ to analyze the
bounce. Strictly speaking, $G_1^{pp}$ and $G_1^{J\bar{J}}$ are also
necessary to implement the reality condition, but we can avoid
explicit solutions since we already verified that the reality
condition is preserved under evolution generated by our truncated
quantum Hamiltonian. The evolution of $G_1^{J\bar{J}}-G_1^{pp}$,
however, will be considered below to analyze effects of the reality
condition on the bounce.

For the real part of $J$ which couples to $p$, the equations give
\begin{eqnarray}
\frac{1}{2}(J+\bar{J})^. &=& -(p+\hbar/2)+
\frac{3}{2}p^2V(\phi)+2\frac{p^3(p+\hbar/2)}{(J-\bar{J})^2}V(\phi)\\
 &&+ \left(\frac{3p^2}{2(J-\bar{J})^2}+
\frac{6p^3(p+\hbar/2)}{(J-\bar{J})^4}\right)
(G^{JJ}+G^{\bar{J}\bar{J}}-2G^{J\bar{J}})V(\phi)\nonumber\\
 &&-3 \left(\frac{p}{J-\bar{J}}+
\frac{4p^2(p+\hbar/2)}{(J-\bar{J})^3}\right) (G^{pJ}-G^{p\bar{J}})V(\phi)
\nonumber\\
&&+\frac{3}{2}\left(1+\frac{4p(p+\hbar/2)}{(J-\bar{J})^2}\right) G^{pp}V(\phi)
\nonumber\\
&&+\frac{p}{(J-\bar{J})^3}(3(J-\bar{J})^2-4p^2)(G^{pJ}-G^{p\bar{J}})V(\phi)
\nonumber\\
&&-\frac{3}{2}\frac{p^2}{(J-\bar{J})^2}
(G^{JJ}+G^{\bar{J}\bar{J}}-2G^{J\bar{J}}-4G^{pp})V(\phi)\,. \nonumber
\end{eqnarray}

Zeroth order solutions (\ref{FreeLoopI})--(\ref{FreeLoopVI}) for this
system now describe bouncing cosmologies. They satisfy
$J_0-\bar{J}_0=2i H_0$ which is the constant free Hamiltonian, and
$G_0^{JJ}+G_0^{\bar{J}\bar{J}}-2 G_0^{J\bar{J}}= 2(c_2-c_1)$ (with
$c_1-c_2=2(\Delta H_0)^2>0$) which can be used to simplify some
calculations.  Perturbed equations of motion are more complicated than
for the Wheeler--DeWitt model, but we can focus on a few dominant
terms. Moreover, we can assume $A=B$ in (\ref{FreeLoopI}),
(\ref{FreeLoopII}) since we would just need to shift our internal time
otherwise.  We keep the classical perturbation, given by
$p^3 (J-\bar{J})^{-2}V(\phi)$ for $\dot{p}$ and
$\frac{3}{2}p^2V(\phi)+2p^3(p+\hbar/2)(J-\bar{J})^{-2}V(\phi)$
for $\frac{1}{2}(J+\bar{J})^{.}$, which contributes (considering only
the highest positive and negative exponents)
\[
 -\frac{J_0+\bar{J}_0}{4H_0^2}p_0^3\sim- A^4\frac{e^{-4\phi}-
 e^{4\phi}}{32H_0^2}
\]
to $\dot{p}_1$ and
\[
 -\frac{p_0^2}{8H_0^2}(4p_0^2-3H_0^2)\sim
-A^4\frac{e^{-4\phi}+e^{4\phi}}{32H_0^2}
\]
to $\frac{1}{2}(J_1+\bar{J}_1)^.$.
This is to be compared with the quantum correction term having the
strongest time dependence, which turns out to be
\[
 \frac{1}{16}\frac{A^2}{H_0^2}
\left(\left(\frac{3}{4}\frac{A^2}{H_0^2} (c_2-c_1)-6c_3+
4\frac{A}{H_0}c_6\right) e^{-4\phi}- \left(\frac{3}{4}\frac{A^2}{H_0^2} (c_2-c_1)-6c_4+
4\frac{A}{H_0}c_5\right) e^{4\phi}\right)
\]
for $\dot{p}$ and
\[
 \frac{1}{16}\frac{A^2}{H_0^2}
\left(\left(\frac{3}{4}\frac{A^2}{H_0^2} (c_2-c_1)-6c_3+
4\frac{A}{H_0}c_6\right) e^{-4\phi}+ \left(\frac{3}{4}\frac{A^2}{H_0^2} (c_2-c_1)-6c_4+
4\frac{A}{H_0}c_5\right) e^{4\phi}\right)
\]
for $\frac{1}{2}(J+\bar{J})^.$.
The inhomogeneous linear differential equations for the first order
perturbations are solved by 
\begin{eqnarray*}
 p_1(\phi) &=& \frac{1}{2}(A_1(\phi)e^{-\phi}+B_1(\phi)e^{\phi})\\
 \frac{1}{2}(J_1(\phi)+\bar{J}_1(\phi)) &=& 
\frac{1}{2}(A_1(\phi)e^{-\phi}-B_1(\phi)e^{\phi})
\end{eqnarray*}
with
\begin{eqnarray*}
 \dot{A}_1(\phi) &=& -\frac{A^2}{16H_0^2}{\cal V}(\phi)
\left(A^2-\frac{3}{2}\frac{A^2}{H_0^2}(c_2-c_1)+12c_3-
8\frac{A}{H_0}c_6\right) e^{-3\phi}\\
 \dot{B}_1(\phi) &=& -\frac{A^2}{16H_0^2}{\cal V}(\phi)
\left(A^2-\frac{3}{2}\frac{A^2}{H_0^2}(c_2-c_1)+12c_4-
8\frac{A}{H_0}c_5\right) e^{3\phi}
\end{eqnarray*}
for the perturbative potential again defined as $V(\phi)=\epsilon{\cal
V}(\phi)$. Thus,
\begin{eqnarray}
 p_1(\phi) &=&
\frac{A^4}{32H_0^2}\left(1-\frac{3}{2}\frac{c_2-c_1}{H_0^2}\right)
\left(e^{\phi}\int^{\phi} {\cal V}(y)e^{3y}\md y- e^{-\phi}\int^{\phi}
{\cal V}(y)e^{-3y}\md y\right)\nonumber\\
&& +\frac{3}{8}\frac{A^2}{H_0^2} \alpha_1 \left(e^{\phi-\delta_1}\int^{\phi} {\cal V}(y)e^{3y}\md y- e^{-\phi+\delta_1}\int^{\phi}
{\cal V}(y)e^{-3y}\md y\right)\\
&&-\frac{1}{4}\frac{A^2}{H_0^2} \alpha_2 \left(e^{\phi-\delta_2}\int^{\phi} {\cal V}(y)e^{3y}\md y- e^{-\phi+\delta_2}\int^{\phi}
{\cal V}(y)e^{-3y}\md y\right)\nonumber
\end{eqnarray}
with 
\begin{equation}
 \alpha_1=\sqrt{c_3c_4} \quad,\quad \alpha_2=\sqrt{c_5c_6} \quad,\quad
 e^{2\delta_1}= \frac{c_3}{c_4} \quad,\quad e^{2\delta_2}=
\frac{c_6}{c_5}\,.
\end{equation}
(Integration constants from integrating $A_1(\phi)$ and $B_1(\phi)$
can be absorbed in the zeroth order ones.)  The parameters $\delta_I$ are
related to the squeezing of the unperturbed state, with $\delta_I=0$ for
an unsqueezed state \cite{BounceCohStates}.

To this perturbative order, the bounce persists even under inclusion
of a matter potential. As in the free model, the key to deciding
whether or not there is a bounce, i.e.\ whether or not $p$ is bounded
away from zero, is the reality condition (\ref{reality}). In the free
case, its implementation resulted in selecting the cosh-solution for
$p$ and ruling out the sinh solution, proving the bounce under the
assumption of a semiclassical state at large volume. (The integration
constants $A$ and $B$ must have equal sign in (\ref{FreeLoopI}) if
$c_1$ is not too large and negative.) Perturbatively, the reality
condition becomes
\begin{eqnarray}
  \frac{1}{4}\hbar^2 &=& |J|^2-(p+\hbar/2)^2-G^{pp}+G^{J\bar{J}}
  = |J_0|^2-(p_0+\hbar/2)^2+ c_1\\
 &&+ 2\epsilon\left({\rm Re}J_0 {\rm Re}J_1+
  {\rm Im}J_0{\rm Im}J_1- (p_0+\hbar/2)p_1
  -{\textstyle\frac{1}{2}}G_1^{pp}
  +{\textstyle\frac{1}{2}}G_1^{J\bar{J}}\right)+ O(\epsilon^2)\,.\nonumber
\end{eqnarray}
The zeroth order is already implemented by starting with the correct
zeroth order solutions. The linear order can then be seen to be automatically
satisfied as well, such that the bouncing behavior at zeroth order is
not affected by the perturbation.

It is, however, also clear that the bounce, with its mixture of
collapsing and expanding functions, leads to stronger quantum
back-reaction than classically. Although $\alpha_1$ and $\alpha_2$ are
small compared to $H_0^2$ for a state which starts out
semiclassically, depending on the values of $\delta_1$ and $\delta_2$
there can be exponential magnifications. For $c_3\approx c_4$ and
$c_5\approx c_6$, which corresponds to nearly unsqueezed states and
implies fluctuations nearly symmetric around the bounce point
\cite{BounceCohStates} quantum corrections are still small at the
bounce if they were small for an initial solution. But for squeezed
states with $c_3$ differing from $c_4$ corrections can be noticeable
even earlier, before the bounce. This is illustrated by the solution
\begin{eqnarray}
 G_1^{J\bar{J}}-G_1^{pp} &=& \frac{1}{36} \frac{A^3}{H_0^2}(c_1-c_2)
\left((2+9\phi^2)\cosh(3\phi)-6\phi\sinh(3\phi)\right)\\
&& -\frac{1}{108}
\frac{A^4}{H_0^3}(c_5-c_6)
\left((2+9\phi^2)\sinh(3\phi)-6\phi\cosh(3\phi)\right)\nonumber\\
&&-\frac{1}{36}\frac{A^3}{H_0^2}\left(4\frac{H_0^2}{A^2}-1\right)
\alpha_1 \left((2+9\phi^2)\cosh(3\phi-\delta_1)-
6\phi\sinh(3\phi-\delta_1)\right)\nonumber\\
&& +\frac{1}{12}\frac{A^2}{H_0}\alpha_2
\left((2+9\phi^2)\cosh(3\phi-\delta_2)-
6\phi\sinh(3\phi-\delta_2)\right)
\nonumber
\end{eqnarray}
for the combination of fluctuations featuring in the reality condition
(specializing to a quadratic potential ${\cal
V}(\phi)=\phi^2$). Especially the sign of this combination is
important since a large negative value can push the bounce in the deep
quantum regime even for large $H$. We have $c_1-c_2=2(\Delta H_0)^2>0$
for the energy fluctuation of free solutions. If we use $A\approx H_0$
which, as a consequence of the reality condition, is satisfied for
free solutions with large $H_0$ and assume unsqueezed states
$c_3\approx c_4$, $c_5\approx c_6$ {\em and} put the extra condition
$c_3\approx c_5$, the remaining terms cancel and
$G_1^{J\bar{J}}-G_1^{pp}$ is positive. But for squeezed states there
can be significant contributions from fluctuations at the bounce,
which can even make $G_1^{J\bar{J}}-G_1^{pp}$ negative there. In such
a situation, it is important to know the precise state, i.e.\ all
parameters $c_1$, $c_2$, $\alpha_1$, $\alpha_2$, $\delta_1$ and
$\delta_2$, in order to determine the quantum nature of the
bounce. However, the relevant squeezing parameters cannot be fully
determined from using the state at only one side of the bounce
\cite{BeforeBB} due to exponential suppression factors of some of the
integration constants in (\ref{FreeLoopIII})--(\ref{FreeLoopVI}) if
one restricts $\phi$ to a fixed sign. Thus, the precise quantum nature
of the bounce may always depend on what assumptions one makes for a
state.

\section{Conclusions}

Loop quantum cosmology implements the discrete nature of quantum
gravity, in contrast to Wheeler--DeWitt quantum cosmology. Its
dynamics is determined by a difference, rather than differential
equation, resulting in singularity-free solutions by general arguments
in symmetric models \cite{Sing,SphSymmSing}. But in general no
intuitive geometrical picture for the avoidance of the classical big
bang singularity is provided; typically the regime around a classical
singularity is of a highly quantum nature. Nevertheless, such regimes
may be probed by effective equations, a first example being presented
in
\cite{QuantumBigBang}. Surprisingly, a very smooth bounce picture
resulted in the model studied there (an isotropic cosmology sourced by
a free, massless scalar), with the whole history described well by
simply replacing a connection component $c$ in the Hamiltonian
constraint by $\sin c$. This replacement is motivated by the use of
holonomies, rather than connection components themselves, in loop
quantum gravity. But there are undoubtedly other corrections in a
general quantum system as well as other types of corrections expected
from a loop quantization, and it was unexpected that only the holonomy
replacement provides effective equations in a precise manner. These
equations were indeed shown to be effective equations in a strict
sense \cite{BouncePert} since other quantum corrections in this
specific model are simply absent. But this analysis also showed that
the validity of effective equations obtained by the holonomy
replacement is very special and tied to the free scalar model.

In this paper, we have provided a detailed systematic analysis for
massive or self-interacting scalars. The situation turns out to be
remarkably different from the free scalar models: back-reaction of
fluctuations on expectation values cannot be ignored, giving rise to
extra quantum corrections to the classical equations. Moreover,
standard as well as more involved choices of setting up an adiabatic
approximation of fluctuations were unsuccessful in that the
adiabaticity assumption was in conflict either with uncertainty
relations or did not lead to any complete solution to all equations of
motion. Thus, for the cases analyzed here there is no analog of a
low-energy effective action. There may be regimes where an
adiabaticity assumption is consistent, but this had to be rather
contrived. Our conclusion is that in general only higher dimensional
effective systems exist for quantum cosmology, which crucially contain
independent quantum degrees of freedom. In particular, the holonomy
replacement of $c$ by $\sin c$ in loop quantum cosmology is not
sufficient, although it still occurs due to the use of $J=pe^{ic}$ as
basic variable in the setup of effective equations.

In general, the holonomy replacement by itself overlooks crucial
quantum effects caused by back-reactions of a spreading and deforming
state on its expectation values. (This can be seen as a crude way of
writing effective equations by setting $G^{a,n}=0$, which obviously
violates uncertainty relations.)  This is important to realize since
the replacement has often occurred in the recent literature, see e.g.\
\cite{RSLoopDual,svv,PhantomLoop,InflRhoSquared,Cosmography}. It is
used either in the form of $\sin c$ in the Friedmann equation, or by
using equations of motion to solve $\sin c$ in terms of $\dot{a}$,
which gives rise to a correction term of energy density squared to the
classical Friedmann equation \cite{RSLoopDual}: starting from a
Hamiltonian constraint of the form $C=-\mu^{-2}\sin^2(\mu
c)\sqrt{|p|}+|p|^{3/2}\rho_{\rm matter}$ (where $\mu$ may be a
$p$-dependent function) one derives the Hamiltonian equation of motion
$\dot{p}= 2\mu^{-1}\sin(\mu c)\cos(\mu c)\sqrt{|p|}$. From $|p|=a^2$,
on the other hand, we have $\dot{p}=2a\dot{a}$ such that the
constraint $C=0$ can be expressed as the corrected Friedmann equation
\begin{equation} \label{rhosquared}
\dot{a}^2= \mu^{-2}\sin^2(\mu c)(1-\sin^2(\mu c))= a^2\rho_{\rm
  matter}(1-(\mu a)^2\rho_{\rm matter})
\end{equation}
using $C=0$. This equation looks very suggestive and easily
generalizable to arbitrary matter densities, which in fact has often
been done in the recent past. However, this equation is applicable
only to an exactly isotropic model sourced by a {\em free, massless
scalar}. Even before embarking on a systematic analysis of effective
equations as in this paper, the holonomy replacement $\sin c$ could
not be expressed simply as a $\rho^2$-correction (or possibly higher
power) if anisotropies or inhomogeneities are included. Then, several
independent gravitational variables exist and one cannot simply solve
for a single $\dot{a}$ to bring the gravitational part of the
Friedmann equation in classical form and have only the matter
contribution corrected.

Moreover, our analysis has shown that the holonomy replacement is not
the only quantum correction when matter interactions are allowed for.
(Similar effects happen when anisotropies or inhomogeneities are
included in addition to the complications for solving for $\dot{a}$
mentioned above.) Quantum fluctuation terms are not only to be
included in the effective equations but even to be kept as independent
variables.\footnote{Even if they were included in the effective
potential, matter terms themselves would be $c$-dependent, and
(\ref{rhosquared}) would only be an implicit equation for $\dot{a}$
whose properties would differ from an explicit one with a
$\dot{a}$-independent right hand side in (\ref{rhosquared}).}  The
replacement of $c$ by $\sin c$ by itself does not result in proper
effective equations at all; such equations are rather to be considered
{\em phenomenological} by isolating one quantum effect but ignoring
others. (A similar type of phenomenological equations without proper
effective analysis occurred in \cite{Inflation} as the first such
example in loop quantum cosmology.) Studies of such phenomenological
equations are valuable, but to avoid misunderstanding they should be
clearly designated as ``phenomenological'' rather than ``effective''
unless an analysis of proper quantum aspects has been performed.

Compared to corrections from inverse powers of metric components,
which were utilized in \cite{Inflation} and elsewhere, for higher
order corrections in $c$ as they result from the holonomy replacement
it is more important to keep fluctuation terms. Both are related to
higher curvature corrections (higher powers of a time derivatives of
metric components on the one hand and higher order time derivatives on
the other) as they are usually expected from effective actions. These
corrections are better not to be separated, or else violations of
covariance may easily occur.

Proper effective equations for flat isotropic models sourced by {\em
massive or self-interacting scalars} have been provided here to first
order in $\hbar$, provided by moments of second order. We have seen
higher powers of $c$ playing a role in the variable $J$, which also
enters equations of motion. In addition, there are clearly quantum
degrees of freedom corresponding to independent fluctuations in the
effective equations. This is in agreement with expectations from
higher curvature terms, although not in obvious correspondence. In our
case, we have true independent quantum degrees of freedom even at the
perturbative level, while higher derivatives in effective actions do
not provide new degrees of freedom as independent solutions in a
consistent perturbation scheme.

We have analyzed the effective equations perturbatively, making sure
that reality conditions are respected. The solutions correspond to
moments computed from a physically normalized state. At first
perturbative order, there is no strong back-reaction for a
Wheeler--DeWitt quantization, or for most of a single collapsing or
expanding branch in a loop quantization. But quantum back-reaction
effects are noticeable for squeezed states around the bounce of loop
quantum cosmology due to the influence of the growing branch on the
collapsing one. Since squeezing parameters of a bouncing state are not
determined by properties in one of the two large volume regime only
\cite{BeforeBB}, no general statement about the presence of a
semiclassical bounce is available. There are quantum parameters whose
influence is negligible at large volume but which will become
important near a bounce and determine its fate.  This shows the
caution required for statements about the behavior of a bounce even in
a perturbative regime around the exact bounce model.  While a full
analysis is beyond the scope of this paper, we have provided the
setting based on a proper derivation and analysis of effective
equations.

\section*{Acknowledgements}

This work was supported in part by NSF grant PHY0554771.

\begin{appendix}

\section{Poisson brackets between fluctuations in
 loop quantum cosmology}
\label{App:Brack}

Here we provide more information on the derivation of Poisson brackets
(\ref{BracketppJbarJ}) -- (\ref{BracketpJpbarJ}). All derivations
follow the same scheme, except for (\ref{BracketppJbarJ}) which does
not give rise to third order moments because
$[\hat{J}\hat{J}^{\dagger},\hat{p}^2]=0=
[\hat{J}^{\dagger}\hat{J},\hat{p}^2]$. In this case, we have
\begin{eqnarray*}
 \{G^{pp},G^{J\bar{J}}\} &=&
\frac{1}{2}\{\langle\hat{p}^2-p^2\rangle,\langle\hat{J}\hat{J}^{\dagger}+
\hat{J}^{\dagger}\hat{J}-2J\bar{J}\rangle\}\\
&=& \frac{1}{2i\hbar} \langle[\hat{p}^2,\hat{J}\hat{J}^{\dagger}+
\hat{J}^{\dagger}\hat{J}]\rangle- \frac{1}{i\hbar} J\langle
[\hat{p}^2,\hat{J}^{\dagger}]\rangle-
\frac{1}{i\hbar}\bar{J}\langle[\hat{p}^2,\hat{J}]\rangle-
\frac{1}{i\hbar}p\langle[\hat{p},\hat{J}\hat{J}^{\dagger}+
\hat{J}^{\dagger}\hat{J}]\rangle\\
&& +\{p^2,J\bar{J}\}\\
&=& -iJ\langle\hat{p}\hat{J}^{\dagger}+\hat{J}^{\dagger}\hat{p}\rangle+
i\bar{J}\langle\hat{p}\hat{J}+\hat{J}\hat{p}\rangle=
-2i(JG^{p\bar{J}}- \bar{J}G^{pJ})\,.
\end{eqnarray*}
For the remaining brackets there are contributions by moments of third
order such as
\begin{eqnarray}
 G^{pJJ} &:=& \frac{1}{3}\langle(\hat{p}-p)(\hat{J}-J)^2+
(\hat{J}-J)(\hat{p}-p)(\hat{J}-J)+ (\hat{J}-J)^2(\hat{p}-p)\rangle\nonumber\\
 &=& \frac{1}{3}\langle\hat{p}\hat{J}^2+\hat{J}\hat{p}\hat{J}+
\hat{J}^2\hat{p}- 3p\hat{J}^2- 3J(\hat{p}\hat{J}+\hat{J}\hat{p})+
6pJ\hat{J}+ 3J^2\hat{p}-3pJ^2\rangle\nonumber\\
&=& \frac{1}{3}\langle\hat{p}\hat{J}^2+\hat{J}\hat{p}\hat{J}+
\hat{J}^2\hat{p}\rangle- 2J(G^{pJ}+pJ)- p(G^{JJ}+J^2)+2pJ^2\nonumber\\
&=& \frac{1}{2}\langle\hat{p}\hat{J}^2+\hat{J}^2\hat{p}\rangle
-2JG^{pJ}- pG^{JJ}-pJ^2 \label{GpJJ}
\end{eqnarray}
where we used $\hat{J}\hat{p}\hat{J}=
\frac{1}{2}(\hat{p}\hat{J}^2+\hat{J}^2\hat{p})$. We will also use the
third moment
\begin{eqnarray}
 G^{ppJ} &=& \frac{1}{3}\langle\hat{p}^2\hat{J}+\hat{p}\hat{J}\hat{p}+
\hat{J}\hat{p}^2\rangle -2p(G^{pJ}+pJ)-J(G^{pp}+p^2)+2p^2J\label{GppJ}
\end{eqnarray}
obtained by a similar calculation, or just by exchanging $p$ and $J$
in $G^{pJJ}$. (But note that the final step in (\ref{GpJJ}) does differ
since we now have
\begin{equation} \label{pJp}
\hat{p}\hat{J}\hat{p}=
\frac{1}{2}(\hat{p}(\hat{p}\hat{J}-\hbar\hat{J})
+(\hat{J}\hat{p}+\hbar\hat{J})\hat{p})=
\frac{1}{2}(\hat{p}^2\hat{J}+\hat{J}\hat{p}^2-\hbar[\hat{p},\hat{J}])=
\frac{1}{2}(\hat{p}^2\hat{J}+\hat{J}\hat{p}^2-\hbar^2\hat{J})\,.
\end{equation}

This allows us to express the Poisson brackets between fluctuations in
terms of moments and expectation values:
\begin{eqnarray*}
 \{G^{pp},G^{JJ}\} &=&
\frac{1}{i\hbar}\langle[\hat{p}^2,\hat{J}^2]\rangle-
\frac{2}{i\hbar}J\langle[\hat{p}^2,\hat{J}]\rangle-
\frac{2}{i\hbar}p\langle[\hat{p},\hat{J}^2]\rangle+ \{p^2,J^2\}\\
&=& -2i\langle\hat{p}\hat{J}^2 +\hat{J}^2\hat{p}\rangle
+2iJ\langle\hat{p}\hat{J}+\hat{J}\hat{p}\rangle
+4ip\langle\hat{J}^2\rangle -4ipJ^2\\
&=& -4iG^{pJJ}-4iJG^{pJ}
\end{eqnarray*}
using (\ref{GpJJ}) in the last step. Furthermore, we obtain
\begin{eqnarray*}
 \{G^{pp},G^{pJ}\} &=&
\frac{1}{2i\hbar}\langle[\hat{p}^2,\hat{p}\hat{J}+\hat{J}\hat{p}]\rangle-
\frac{1}{i\hbar}p\langle[\hat{p}^2,\hat{J}]\rangle-
\frac{1}{i\hbar}p\langle[\hat{p},\hat{p}\hat{J}+\hat{J}\hat{p}]\rangle+
\{p^2,pJ\}\\
&=& -\frac{1}{2}i\langle\hat{p}^2\hat{J}+2\hat{p}\hat{J}\hat{p}+
\hat{J}\hat{p}^2\rangle +2ip\langle\hat{p}\hat{J}+\hat{J}\hat{p}\rangle
-2ip^2J\\
&=& -\frac{2}{3}i\langle\hat{p}^2\hat{J}+\hat{p}\hat{J}\hat{p}+
\hat{J}\hat{p}^2\rangle+ \frac{1}{6}i\hbar^2J +4ip(G^{pJ}+pJ) -2ip^2J\\
&=&-2iG^{ppJ}-2iJG^{pp}+\frac{1}{6}i\hbar^2J
\end{eqnarray*}
where we used (\ref{pJp}) to bring the third moment in the form as it
appears in (\ref{GppJ}).  The remaining brackets follow from similar
calculations, although they can become more lengthy if all three basic
operators $\hat{p}$, $\hat{J}$ and $\hat{J}^{\dagger}$ are involved.
Skipping some of the details of calculational steps already
encountered, and occasionally using the commutation relations as well
as $\hat{J}\hat{J}^{\dagger}=\hat{p}^2$ and the associated reality
condition (\ref{reality}), we further have
\begin{eqnarray*}
 \{G^{J\bar{J}},G^{JJ}\} &=& \frac{1}{2i\hbar}
 \langle[\hat{J}\hat{J}^{\dagger}+
 \hat{J}^{\dagger}\hat{J},\hat{J}^2]\rangle
 -\frac{1}{i\hbar}J\langle[\hat{J}\hat{J}^{\dagger}+
 \hat{J}^{\dagger}\hat{J},\hat{J}]\rangle
 -\frac{1}{i\hbar}J\langle[\hat{J}^{\dagger},\hat{J}^2]\rangle
 +\{J\bar{J},J^2\} \\
&=& -i\langle\hat{J}^2\hat{p} +2\hat{J}\hat{p}\hat{J}
+\hat{p}\hat{J}^2+2\hbar\hat{J}^2\rangle
+4iJ\langle\hat{p}\hat{J}+\hat{J}\hat{p}\rangle -4ipJ^2+2i\hbar J^2\\
&=& -4iG^{pJJ} -8iJ(G^{pJ}+pJ) -4ip(G^{JJ}+JJ) +8ipJ^2
-2i\hbar(G^{JJ}+J^2)\\
&& +8iJ(G^{pJ}+pJ) -4ipJ^2+2i\hbar J^2
= -4iG^{pJJ}-4i(p+\hbar/2)G^{JJ}\,,
\end{eqnarray*}
\begin{eqnarray*}
 \{G^{J\bar{J}},G^{pJ}\} &=& \frac{1}{4i\hbar}
 \langle[\hat{J}\hat{J}^{\dagger}+
 \hat{J}^{\dagger}\hat{J},\hat{p}\hat{J}+\hat{J}\hat{p}]\rangle
 -\frac{1}{2i\hbar}p\langle[\hat{J}\hat{J}^{\dagger}+
 \hat{J}^{\dagger}\hat{J},\hat{J}]\rangle
 -\frac{1}{2i\hbar}J\langle[\hat{J}\hat{J}^{\dagger}+
 \hat{J}^{\dagger}\hat{J},\hat{p}]\rangle\\
&& -\frac{1}{2i\hbar}J\langle[\hat{J}^{\dagger},\hat{p}\hat{J}+
\hat{J}\hat{p}]\rangle
 -\frac{1}{2i\hbar}\bar{J}\langle[\hat{J},\hat{p}\hat{J}+
\hat{J}\hat{p}]\rangle
 +\{J\bar{J},pJ\} \\
&=& -\frac{1}{2}i\langle\hat{p}^2\hat{J}+2\hat{p}\hat{J}\hat{p}
+\hat{J}\hat{p}^2\rangle
+i(p+\hbar)\langle\hat{p}\hat{J}+\hat{J}\hat{p}\rangle +2i\hbar pJ
+2iJ\langle\hat{p}^2\rangle\\
&&+\frac{1}{2}iJ\langle\hat{J}\hat{J}^{\dagger}
+\hat{J}^{\dagger}\hat{J}\rangle -i\bar{J}\langle\hat{J}^2\rangle
-2ip(p+\hbar/2)J\\
&=& -2iG^{ppJ}-2i(p+\hbar/2)G^{pJ} +iJG^{J\bar{J}} -i\bar{J}G^{JJ}
+\frac{1}{6}i\hbar^2J\,,
\end{eqnarray*}
\begin{eqnarray*}
 \{G^{JJ},G^{pJ}\} &=& \frac{1}{2i\hbar}
 \langle[\hat{J}^2,\hat{p}\hat{J}+\hat{J}\hat{p}]\rangle
 -\frac{1}{i\hbar}J\langle[\hat{J}^2,\hat{p}]\rangle
 -\frac{1}{i\hbar}J\langle[\hat{J},\hat{p}\hat{J}+\hat{J}\hat{p}]\rangle
 +\{J^2,pJ\} \\
&=& 2i\langle\hat{J}^3\rangle -4iJ\langle\hat{J}^2\rangle +2iJ^3 =
2iG^{JJJ} +2iJG^{JJ}\,,
\end{eqnarray*}
\begin{eqnarray*}
 \{G^{JJ},G^{p\bar{J}}\} &=& \frac{1}{2i\hbar}
 \langle[\hat{J}^2,\hat{p}\hat{J}^{\dagger}+\hat{J}^{\dagger}\hat{p}]\rangle
 -\frac{1}{i\hbar}\bar{J}\langle[\hat{J}^2,\hat{p}]\rangle
 -\frac{1}{i\hbar}J\langle[\hat{J},\hat{p}\hat{J}^{\dagger}+
\hat{J}^{\dagger}\hat{p}]\rangle +\{J^2,p\bar{J}\} \\
&=& 2i\langle\hat{p}^2\hat{J} +\hat{p}\hat{J}\hat{p}+\hat{J}\hat{p}^2
+2\hbar(\hat{p}\hat{J}+\hat{J}\hat{p})\rangle
-2ip\langle\hat{p}\hat{J}+\hat{J}\hat{p}+\hbar\hat{J}\rangle
-2i\bar{J}\langle\hat{J}^2\rangle\\
&&-iJ\langle\hat{J}\hat{J}^{\dagger}+\hat{J}^{\dagger}\hat{J}
+4\hat{p}(\hat{p}+\hbar/2)\rangle +2iJ^2\bar{J} +4ip(p+\hbar/2)J\\
&=& 6iG^{ppJ}+8i(p+\hbar/2)G^{pJ} +2iJG^{pp} -2i\bar{J}G^{JJ}
-2iJG^{J\bar{J}} +2ip^2J -2iJ^2\bar{J}\\
&& +2i\hbar pJ\\
&=& 6iG^{ppJ} +8i(p+\hbar/2)G^{pJ} -2i\bar{J}G^{JJ}-i\hbar^2J\,,
\end{eqnarray*}
\begin{eqnarray*}
 \{G^{pJ},G^{p\bar{J}}\} &=& \frac{1}{4i\hbar}
 \langle[\hat{p}\hat{J}+
 \hat{J}\hat{p},\hat{p}\hat{J}^{\dagger}+\hat{J}^{\dagger}\hat{p}]\rangle
 -\frac{1}{2i\hbar}p\langle[\hat{J},\hat{p}\hat{J}^{\dagger}+
 \hat{J}^{\dagger}\hat{p}]\rangle
 -\frac{1}{2i\hbar}J\langle[\hat{p},\hat{p}\hat{J}^{\dagger}+
 \hat{J}^{\dagger}\hat{p}]\rangle\\
&& -\frac{1}{2i\hbar}p\langle[\hat{p}\hat{J}+
\hat{J}\hat{p},\hat{J}^{\dagger}]\rangle
 -\frac{1}{2i\hbar}\bar{J}\langle[\hat{p}\hat{J}+
\hat{J}\hat{p},\hat{p}]\rangle
 +\{pJ,p\bar{J}\} \\
&=& \frac{1}{4}i\langle8\hat{p}^3
+\hat{p}(\hat{J}\hat{J}^{\dagger}+2\hat{J}^{\dagger}\hat{J})
+(\hat{J}\hat{J}^{\dagger}+2\hat{J}^{\dagger}\hat{J})\hat{p}
+\hat{J}\hat{p}\hat{J}^{\dagger} +\hat{J}^{\dagger}\hat{p}\hat{J}
+4\hbar\hat{p}^2\rangle\\
&& -ip\langle\hat{J}\hat{J}^{\dagger}
+\hat{J}^{\dagger}\hat{J} +4\hat{p}^2+2\hbar\hat{p}\rangle
-\frac{1}{2}iJ\langle\hat{p}\hat{J}^{\dagger}+\hat{J}^{\dagger}\hat{p}\rangle
-\frac{1}{2}i\bar{J}\langle\hat{p}\hat{J}+\hat{J}\hat{p}\rangle\\
&&+2ipJ\bar{J} +2ip^2(p+\hbar/2)\\
&=& 4iG^{ppp}+i(8p+3\hbar)G^{pp} -2ipG^{J\bar{J}} -iJG^{p\bar{J}}
-i\bar{J}G^{pJ} +2ip^3-2ipJ\bar{J}\\
&&+2i\hbar p^2+\frac{3}{2}i\hbar^2p+\frac{1}{4}i\hbar^3\\
&=& 4iG^{ppp} +6i(p+\hbar/2)G^{pp}-iJG^{p\bar{J}} -i\bar{J}G^{pJ}
+\frac{1}{2}i\hbar^2p+\frac{1}{4}i\hbar^3
\end{eqnarray*}
and
\begin{eqnarray*}
 \{G^{JJ},G^{\bar{J}\bar{J}}\} &=&
   \frac{1}{i\hbar}\langle[\hat{J}^2,(\hat{J}^{\dagger})^2]\rangle
   -\frac{2}{i\hbar}J\langle[\hat{J},(\hat{J}^{\dagger})^2]\rangle
   -\frac{2}{i\hbar}\bar{J}\langle[\hat{J}^2,\hat{J}^{\dagger}]\rangle
   +\{J^2,\bar{J}^2\}\\
&=& 2i\langle\hat{J}\hat{p}\hat{J}^{\dagger}
+\hat{J}\hat{J}^{\dagger}\hat{p} +\hat{p}\hat{J}^{\dagger}\hat{J}
+\hat{J}^{\dagger}\hat{p}\hat{J}
+\hbar(\hat{J}\hat{J}^{\dagger}+\hat{J}^{\dagger}\hat{J})\rangle
-4iJ\langle\hat{J}^{\dagger}\hat{p}+\hat{p}\hat{J}^{\dagger}+
\hbar\hat{J}^{\dagger}\rangle\\
&&-4i\bar{J}\langle\hat{J}\hat{p}+\hat{p}\hat{J} +\hbar\hat{J}\rangle
+8i(p+\hbar/2)J\bar{J}\\
&=& 8iG^{ppp}+16i(p+\hbar/2)G^{pp} +8i(p+\hbar/2)G^{J\bar{J}}
-8iJG^{p\bar{J}}\\
&& -8i\bar{J}G^{pJ} +4i\hbar^2p+2i\hbar^3\,.
\end{eqnarray*}

\end{appendix}


\end{document}